\documentclass[11pt]{article} % use larger type; default would be 10pt

\usepackage{amssymb}
\usepackage{array}
\usepackage{fancyhdr}
\usepackage{subfigure}
\usepackage{graphicx}
\usepackage{mathrsfs}
\usepackage{slashed}
\usepackage{stmaryrd}
\usepackage{epsfig}
\usepackage{dcolumn}
\usepackage{bm}
\usepackage{amsmath}
\usepackage{cite}
\usepackage{simplewick}
 \usepackage{relsize}

\usepackage{array}
\usepackage{fancyhdr}
\usepackage{graphicx}
\usepackage{mathrsfs}
\usepackage{slashed}
\usepackage{stmaryrd}

\usepackage{wasysym}

\usepackage{amssymb}
\usepackage{array}
\usepackage{fancyhdr}
\usepackage{subfigure}
\usepackage{graphicx}
\usepackage{mathrsfs}
\usepackage{slashed}
\usepackage{stmaryrd}
\usepackage{epsfig}
\usepackage{dcolumn}
\usepackage{bm}
\usepackage{amsmath}
\usepackage{cite}

\usepackage[utf8]{inputenc} % set input encoding (not needed with XeLaTeX)

\usepackage{geometry} % to change the page dimensions
\usepackage{graphicx} % support the \includegraphics command and options

\usepackage{booktabs} % for much better looking tables
\usepackage{array} % for better arrays (eg matrices) in maths
\usepackage{paralist} % very flexible & customisable lists (eg. enumerate/itemize, etc.)
\usepackage{verbatim} % adds environment for commenting out blocks of text & for better verbatim
\usepackage{subfig} % make it possible to include more than one captioned figure/table in a single float
\usepackage{mathtools}

\usepackage{fancyhdr} % This should be set AFTER setting up the page geometry
\usepackage{sectsty}
\allsectionsfont{\sffamily\mdseries\upshape} % (See the fntguide.pdf for font help)
\usepackage[nottoc,notlof,notlot]{tocbibind} % Put the bibliography in the ToC
\usepackage[titles,subfigure]{tocloft} % Alter the style of the Table of Contents

\title{An Electroweak-like Theory from Four Fermion Interactions}
\author{Yi-Cheng Huang}
\begin{document}
\maketitle
%%%%%%%%%%%%%%%%%%%%%%%%%%%%%%%%%%%%%%%%%%%%%%%%%%%%%%%%%%%%%%%%%%%%%%%%%%%%%%%%%%%
%%%%%%%%%%%%%%%%%%%%%%%%%%%%%%%%%%%%%%%%%%%%%%%%%%%%%%%%%%%%%%%%%%%%%%%%%%%%%%%%%%%
\abstract
An electroweak-like  theory of a broken chiral symmetry that is constructed by the collective modes of fermion pairs from four fermion interactions of one lepton generation is presented. The products of Dirac spinors lead to the separation of the two chiral fermions to couple respectively with two different kinds of polarization states.  Because of a broken vacuum, a fermion and an anti-fermion out of the four pair up to form vector bosons, which behave like gauge bosons, such as $W^\pm$, $Z$ and $\gamma$ in a group structure of $SU(2)_L\times U(1)_Y$. The pairing of spinors only allows left-handed fermions to interact with charged bosons to secure the gauge invariance, while, as desired, $Z$-like bosons mediate different weak forces for two chiral fermions and $\gamma$-like bosons interact freely with fermions.

%\newpage
%%%%%%%%%%%%%%%%%%%%%%%%%%%%%%%%%%%%%%%%%%%%%%%%%%%%%%%%%%%%%%%%%%%%%%%%%%%%%%%%%%%
%%%%%%%%%%%%%%%%%%%%%%%%%%%%%%%%%%%%%%%%%%%%%%%%%%%%%%%%%%%%%%%%%%%%%%%%%%%%%%%%%%%
\section{Introduction}
In the BCS theory \cite{bcs}, which describes the superconductivity \cite{tinkham} in solids, for the electrons near Fermi surface, their instability aggravates and the formation of stable binding pairs occurs at a quite low temperature. The so-called Cooper pairs are attributed to the electron-lattice interaction. The resultant attractive interaction between electrons can be captured by the BCS four-operator interaction Hamiltonian: $-\sum V_0 \,C^\dagger_{{\bf - k}\downarrow}C^\dagger_{{\bf k}\uparrow}C_{-{\bf p}\downarrow}C_{{\bf p}\uparrow}$, and it leads to a gapped ground state of coherent phases of the electron pairs of Block states $({\bf p}\uparrow,-{\bf p}\downarrow)$, which breaks the $U(1)$ symmetry. It is because of this coherent ground state, the expectation values of the pairs, $\langle C_{{\bf -p}\downarrow}C_{{\bf p}\uparrow}\rangle$, becomes nonzero. Thus, in effect a pair of electrons is correlated due to a seemingly attractive interaction between them. Lots of such pairs overlap and form a strong collective condensate, and it becomes harder to break the condensate as a whole. As a result, electron flows propagate and convey the energy without resistance in solids. This is the picture that the BCS theory has given us  since the first superconductor, solid mercury, whose resistance abruptly dropped to zero at the temperature 4.2 K,  was found in the early 20-th century  \cite{onnes11}. On the other hand, the electroweak interaction is also explained by a mathematical formalism with a symmetry breaking, and a field theory \cite{huang13}, which is constructed in the imaginary-time formulation, is able to describe QED as a limiting case of vacuum at zero temperature. It would be of interest to investigate if a similar mechanism could occur in a thermodynamic perspective for weak interactions and make an electroweak theory, which has to be a theory of symmetry breaking. Even though it is not clear at this point how a vacuum is broken from this point of view in order to fit into the celebrated Higgs mechanism \cite{higgs}, it would still be inspiring to see if it is possible to construct electroweak interactions \cite{electroweak} based on the same picture. In this manuscript, such an attempt is made as follows. All possible four-fermion interactions in the form, $\bar{\psi}\bar{\psi}\psi\psi$, are examined and the pairings of the operators are formulated to form four vector bosons for one lepton generation. Before the vacuum is broken, the bosons are massless.  %The Lorentz invariance of the Lagrangian for the imaginary-time and space is used to deduce the polarization vectors.
It is found that the charged bosons in the group of $SU(2)_L\times U(1)_Y$ interact only with left-handed fermions, while those  from $SU(2)_R\times U(1)_Y$ couple with right-handed ones. Both kinds of fermions are able to coexist with the chiral symmetry. In the broken phase of vacuum, the massive bosons of  $SU(2)_R$  become non-physical due to violation of the Ward identity. The neutral bosons could interact with right-handed fermions when either the energy or momentum  of bosons is zero, thus the photons defined in this formalism are free to interact equally with fermions regardless of their chiralities.  %One thing to emphasize is that the whole formalism is discussed under the imaginary-time and space, as well as the conditions of gauge invariance. Presumptively, the Lagrangian is gauge invariant for the imaginary-time, as it should be. 
\par
Traditionally an interaction in field theories is given by requiring the Lagrangian having some symmetries, like gauge, chiral, or both, etc.; in the idea presented here the collective modes of four fermion interaction are considered in the beginning to satisfy the requirements of the given symmetries. In the next section, all possible pairings of Dirac spinors are discussed. Different pairing of their spins gives rise to distinct polarization vectors of a vector boson. In Section 3, a toy model of fermion pairs are then presented to formulate four vector bosons, which appear to assume the roles that are played by $W^\pm$, $Z$ and $\gamma$ bosons, to mediate forces between fermions of the same or different flavors. The formulation of groups of $SU(2)\times U(1)$ from all of four-fermion interactions are presented in the following section. The physical meanings of the theory will be discussed in Conclusion.

%%%%%%%%%%%%%%%%%%%%%%%%%%%%%%%%%%%%%%%%%%%%%%%%%%%%%%%%%%%%%%%%%%%%%%%%%%%%%%%%%%%
%%%%%%%%%%%%%%%%%%%%%%%%%%%%%%%%%%%%%%%%%%%%%%%%%%%%%%%%%%%%%%%%%%%%%%%%%%%%%%%%%%%
\section{Four Fermion Interactions}
Here, a series of four-fermion interactions will be discussed. Two types of fermions, $A$- and $B$-type, are used in the derivations. It will be clear in the following sections that $A$-type fermions play a role of electrons, and $B$-type refers to neutrinos. At this point, a gauge transformation has not been introduced yet, and we may just treat these two types of fermions equally without other presumptions. In the imaginary-time and space, $(\tau,{\bf x})$, the possible forms of the interaction are 
 \begin{eqnarray}
\begin{matrix}
&&(a)\sum\limits_{r,s,r',s'}\overline{\psi}^{\,r'}_{A,\alpha}  \psi^{r}_{A,\alpha} \overline{\psi}^{\,s}_{B,\beta} \psi^{s'}_{B,\beta}(\tau,{\bf x}),\hspace{0.95cm}(b)\sum\limits_{r,s,r',s'}\overline{\psi}^{\,r'}_{A,\alpha}  \psi^{r}_{A,\beta} \overline{\psi}^{\,s}_{B,\beta} \psi^{s'}_{B,\alpha}(\tau,{\bf x}),\\
&&(c)\sum\limits_{r,s,r',s'}\overline{\psi}^{\,r'}_{A,\alpha}  \psi^{r}_{A,\alpha} \overline{\psi}^{\,s}_{A,\beta} \psi^{s'}_{A,\beta}(\tau,{\bf x}),\hspace{.2cm} {\rm and}\hspace{.2cm}(d)\sum\limits_{r,s,r',s'}\overline{\psi}^{\,r'}_{B,\alpha}  \psi^{r}_{B,\beta} \overline{\psi}^{\,s}_{B,\beta} \psi^{s'}_{B,\alpha}(\tau,{\bf x}),
\end{matrix}\label{4fermion}
 \end{eqnarray}
where Greek alphabets, $\alpha$ and $\beta$ in the subscripts, indicate Dirac or spinor indices and English ones, $r$, $s$, etc., are for spin states. Notice if $\beta$ is not a subindex, it means ${\frac{1}{k_B T}}$. In the following, the symbol, the double bracket $\langle \langle ...\rangle\rangle $, will refer to the pairing of two fermions, such as 
  \begin{eqnarray*}
\langle \langle\psi^{r}_{A,\alpha} \overline{\psi}^{\,s}_{A,\beta} \rangle\rangle, \hspace{.5cm}\langle \langle\psi^{r}_{A,\alpha} \overline{\psi}^{\,s}_{B,\beta} \rangle\rangle, \hspace{0.5cm} {\rm etc.}\,.
 \end{eqnarray*}
Throughout this manuscript, the indices of Dirac spinors will be kept in order not to confuse column spinors and their transpose; therefore $\psi^\dagger$ simply means the conjugate of the field operator $\psi$ without a meaning of a transpose once a spin-index is added.   For the imaginary-time and space \cite{huang13}, the fermion field operators are expanded as follows
 \begin{eqnarray*}
\psi^r_{A,\alpha} (x)&=&\frac{1}{\beta}\sum_n \int \frac{d^3 {\bf p}}{(2\pi)^3}\left(a^{r}_{\omega_n,{\bf p}} u_\alpha^r({\bf p}) e^{-i\omega_n\tau-i{\bf p\cdot x}}+c^{r,\dagger}_{\omega_n,{\bf p}} v_\alpha^r({\bf p}) e^{i\omega_n\tau+i{\bf p\cdot x}}\right),
\\%\end{eqnarray*}
 %\begin{eqnarray*}
\psi^r_{B,\alpha} (x)&=&\frac{1}{\beta}\sum_n \int \frac{d^3 {\bf p}}{(2\pi)^3}\left(b^{r}_{\omega_n,{\bf p}} u_\alpha^r({\bf p}) e^{-i\omega_n\tau-i{\bf p\cdot x}}+d^{s,\dagger}_{\omega_n,{\bf p}} v_\alpha^r({\bf p}) e^{i\omega_n\tau+i{\bf p\cdot x}}\right),
\end{eqnarray*}
where $u_\alpha^r({\bf p})$ and $v_\alpha^r({\bf p})$ are the spinors for fermion and anti-fermion. %, and the four momentum $p=(p_0,{\bf p})$, where $p_0=\sqrt{{\bf p}^2+m^2}$.    
 For fermions, the integer $n$ of the Matsubara frequency, $\omega_n\,(=\mathsmaller{\frac{n \pi }{\beta}})$, is an odd number.  In the conventional field expansions a factor $\frac{1}{\sqrt{2\xi_{\bf p}}}$ has to be added to secure the Lorentz invariance, where $\xi_{\bf p}=\sqrt{{\bf p}^2+m^2}$. It will be ignored at this stage, since it is not relevant to the following derivations.  %The annihilation and creation operators obey the anti-commutation relations:
% \begin{eqnarray*}
%\{a^{r}_{\omega_{n},{\bf p}},a^{s\dagger}_{\omega_m,{\bf q}}\}
%=(2\pi)^3\beta^3\delta_{rs}\delta_{mn}\delta^{(3)}({\bf p-q}),&&
%\{b^{r}_{\omega_{n},{\bf p}},b^{s\dagger}_{\omega_m,{\bf q}}\}
%=(2\pi)^3\beta^3\delta_{rs}\delta_{mn}\delta^{(3)}({\bf p-q}),\\
%\{c^{r}_{\omega_{n},{\bf p}},c^{s\dagger}_{\omega_m,{\bf q}}\}
%=(2\pi)^3\beta^3\delta_{rs}\delta_{mn}\delta^{(3)}({\bf p-q}),&&
%\{d^{r}_{\omega_{n},{\bf p}},d^{s\dagger}_{\omega_m,{\bf q}}\}
%=(2\pi)^3\beta^3\delta_{rs}\delta_{mn}\delta^{(3)}({\bf p-q}).
% \end{eqnarray*}

%%%%%%%%%%%%%%%%%%%%%%%%%%%%%%%%%%%%%%%%%%%%%%%%%%%%%%%%%%%%%%%%%%%%%%%%%%%%%%%%%%%
\subsection{Pair-up of spinors}
From the quantum theory \cite{sakurai93}, the addition of angular momentum for two spin one-half particles forms three states of a spin-one and a spin-0 particles.  In terminology, it can be expressed by the notation, $\frac{1}{2}\otimes\frac{1}{2}=1\oplus 0$. Besides, each of three states of the spin-one particle refers to distinctive spatial orientation. For example, the spherical harmonic functions, $Y^m_l(\phi,\theta)$, which may be denoted as $\langle \phi,\theta|l,m\rangle$, for the total angular momentum, $l=1$, are
 \begin{eqnarray}
Y^0_1=\sqrt{\frac{3}{4\pi}}V_z,\hspace{.5cm} {\rm and} \hspace{.5cm} Y^{\pm 1}_1=\sqrt{\frac{3}{4\pi}}
\left(\mp\frac{V_x\pm iV_y}{\sqrt{2}}\right), \label{shf}
 \end{eqnarray}
where $(V_x,V_y,V_z)=(x,y,z)$ is a position vector. Based on the same reason, it will be shown soon that the addition of two spin states forms the four polarization states of a vector boson.  For a simplicity reason and without losing any generality, the solutions of spinors for a fermion and a anti-fermion, which are assumed to propagate only in the $z$-direction with a momentum $p^\mu_n=(p^0,0,0,p^3)$, are given as 
 \begin{eqnarray*} 
u_\alpha^r(p)= 
\begin{pmatrix}
 \sqrt{p\cdot \sigma}  \xi_r \\
 \sqrt{p\cdot \bar{\sigma}}  \xi_r 
\end{pmatrix}, &&
v_\alpha^r(p)= 
\begin{pmatrix}
 \sqrt{p\cdot \sigma}  \eta_r \\
 -\sqrt{p\cdot \bar{\sigma}}  \eta_r 
\end{pmatrix},
 \end{eqnarray*}
where two-component spinors are
 \begin{eqnarray*} 
\xi_{\mathsmaller{\frac{1}{2}}}=
\begin{pmatrix}
1 \\ 0
\end{pmatrix},\hspace{.5cm}
\xi_{\mathsmaller{-\frac{1}{2}}}=
\begin{pmatrix}
 0\\ 1
\end{pmatrix},\hspace{.5cm}
\eta_{\mathsmaller{\frac{1}{2}}}=
\begin{pmatrix}
0 \\ 1
\end{pmatrix},\hspace{.3cm}{\rm and} \hspace{.3cm}
\eta_{\mathsmaller{-\frac{1}{2}}}=
\begin{pmatrix}
 -1\\ 0
\end{pmatrix},
 \end{eqnarray*} and ${\sigma}^\mu=({\bf 1}, \vec{\sigma})$ are a $2\times 2$ identity and  Pauli matrices and $\bar{\sigma}^\mu=({\bf 1}, -\vec{\sigma})$. The combinations for the product of the spinors are computed to obtain
\begin{eqnarray} 
\hspace{-0cm}
\begin{matrix}
u_\alpha^{\mathsmaller{\frac{1}{2}}}(p) v_\beta^{\mathsmaller{\frac{1}{2}}}(p)
-u_\alpha^{\mathsmaller{-\frac{1}{2}}}(p) v_\beta^{\mathsmaller{-\frac{1}{2}}}(p)
= &\gamma^0\gamma^1\left(-m+p^0\gamma^0-p^3\gamma^3\right)\Rightarrow \gamma^0\gamma^1\left(\slashed{p}-m\right),\\
u_\alpha^{\mathsmaller{\frac{1}{2}}}(p) v_\beta^{\mathsmaller{\frac{1}{2}}}(p)
+u_\alpha^{\mathsmaller{-\frac{1}{2}}}(p) v_\beta^{\mathsmaller{-\frac{1}{2}}}(p)
= & i\gamma^0\gamma^2\left(-m+p^0\gamma^0-p^3\gamma^3\right)\Rightarrow i\gamma^0\gamma^2\left(\slashed{p}-m\right),\\
u_\alpha^{\mathsmaller{\frac{1}{2}}}(p) v_\beta^{\mathsmaller{-\frac{1}{2}}}(p)
+u_\alpha^{\mathsmaller{-\frac{1}{2}}}(p) v_\beta^{\mathsmaller{\frac{1}{2}}}(p)
= &-\gamma^3\left(-m+p^0\gamma^0-p^3\gamma^3\right)\Rightarrow -\gamma^3\left(\slashed{p}-m\right),\\
u_\alpha^{\mathsmaller{-\frac{1}{2}}}(p) v_\beta^{\mathsmaller{\frac{1}{2}}}(p)
-u_\alpha^{\mathsmaller{\frac{1}{2}}}(p) v_\beta^{\mathsmaller{-\frac{1}{2}}}(p)
= &\gamma^0\gamma^5\left(-m+p^0\gamma^0-p^3\gamma^3\right)\Rightarrow \gamma^0\gamma^5\left(\slashed{p}-m\right),
\end{matrix}\label{produv}
 \end{eqnarray}   
where $\slashed{p}=p^\mu\gamma_\mu$ for a 4-momentum $p^\mu$. The difference of fermion masses will not taken into account in this toy model, and the expressions in Eq. (\ref{produv}) should be still a good approximation as $|p|\gg m$ even two fermions have different masses. As for this issue of masses, there will be more discussions in Conclusion. From above, the gamma matrices of each formula  imply the the orientation in the four dimensional time and space. 
%where $|p|=\sqrt{p_0^2-{\bf p}^2}$. If the particle is on-shell, $|p|$ equal to its mass. 
The first and second formulas of the above are  analogous to the second one in Eq. (\ref{shf}); the subtraction of the states $|1,1\rangle\,(=u^{\mathsmaller{\frac{1}{2}}} v^{\mathsmaller{\frac{1}{2}}})$ and $|1,-1\rangle\,(=u^{\mathsmaller{-\frac{1}{2}}} v^{\mathsmaller{-\frac{1}{2}}})$ corresponds to $x$-direction, and the addition of the two  refers to $y$-direction. % due to the associated gamma matrices, $\gamma^1$ and $\gamma^2$. 
Notice these are the gamma matrices that will appear in a fermion current, $\overline{\psi}\gamma^\mu \psi$. The state, $|1,0\rangle\,(=u^{\mathsmaller{\frac{1}{2}}} v^{\mathsmaller{-\frac{1}{2}}}+u^{\mathsmaller{-\frac{1}{2}}} v^{\mathsmaller{\frac{1}{2}}})$, is pointing to $z$-direction. The state, $|0,0\rangle\,(=u^{\mathsmaller{-\frac{1}{2}}} v^{\mathsmaller{\frac{1}{2}}}-u^{\mathsmaller{\frac{1}{2}}} v^{\mathsmaller{-\frac{1}{2}}})$, does not belong to any spatial dimension; it will be assigned to the time component as the polarization states of vector bosons are constructed. An example of the four composite spin states mixed together due to Lorentz boosts is the Wigner rotation \cite{wignerrotation}: an inertia effect of the total angular momentum of two paired electrons, which is observed by a moving observer. Another reason is that if the two spin indices are contracted, the state, $|1,0\rangle$, either for the left- or right-handed fermions in the third line of Eq. (\ref{produv}), is  $-2\,p^3$, and $|0,0\rangle$ in the last line  for left- and right-handed fermions are  $2\,p^0$ and $-2\,p^0$. Both of their 4-vectors are Lorentz invariant.   Regardless of $x$- and $y$-dimensions, combine the zero-th and third components as a 2-vector for the particle's rest frame, so that $p^\mu=(m,0,0,0)$:
\begin{eqnarray}
V^a(0)=
\begin{pmatrix}
u^{-\frac{1}{2}}_{\alpha} (0)v^{\frac{1}{2}}_{\beta}(0)-u^{\frac{1}{2}}_{\alpha} (0)v^{-\frac{1}{2}}_{\beta}(0)\\
u^{-\frac{1}{2}}_{\alpha} (0)v^{\frac{1}{2}}_{\beta}(0)+u^{\frac{1}{2}}_{\alpha} (0)v^{-\frac{1}{2}}_{\beta}(0)
\end{pmatrix},\label{Va}
 \end{eqnarray}
where the index $a=0,3$, for the upper and the lower component of the 2-vector and $u^r(0)$ and $v^r(0)$ are the spinors in parrticle's rest frame.  In order to obtain the 2-vector by an observer, it takes a Lorentz boost to transform from $u^r(0)$ to $u^r(p)$, so does $v^r(p)$. However, the above 2-vector, $V^{a}(p)$, after a boost for the spinors, is still in particle's rest frame.  Therefore what needs to be known is how this 2-vector is described in the observer's reference frame and again we know it simply takes a Lorentz transformation. After a boost is applied on $V^a(p)$ with a 4-momentum ${p}^\mu=(p^0,0,0,p^3)$, the 2-vector of the observer's frame, $V^{a'}(p)$, is
  \begin{eqnarray}
V^{a'}(p)&=&\frac{1}{|p|}
\begin{pmatrix}
p_0&p_3\\
p_3&p_0
\end{pmatrix}V^a,\nonumber
\\
&=&\frac{1}{|p|}
\begin{pmatrix}
(p_0+p_3) u^{\mathsmaller{-\frac{1}{2}}}_{\alpha} (p)v^{\mathsmaller{\frac{1}{2}}}_{\beta}(p)-(p_0-p_3)u^{\mathsmaller{\frac{1}{2}}}_{\alpha} (p)v^{\mathsmaller{-\frac{1}{2}}}_{\beta}(p)\\
(p_0+p_3)u^{\mathsmaller{-\frac{1}{2}}}_{\alpha} (p)v^{\mathsmaller{\frac{1}{2}}}_{\beta}(p)+(p_0-p_3)u^{\mathsmaller{\frac{1}{2}}}_{\alpha} (p)v^{\mathsmaller{-\frac{1}{2}}}_{\beta}(p)
\end{pmatrix}\label{Vb},
 \end{eqnarray}
where ${p}_\mu=(p^0,0,0,-p^3)$ in a conventional notation of tensors and $|p|=\sqrt{p^\mu p_\mu}$. While comparing Eq. (\ref{Va}) with Eq. (\ref{Vb}), it can be concluded that, for each of the two spinor products, when they are boosted, 
 \begin{eqnarray}
u^{-\frac{1}{2}}_{\alpha} (0)v^{\frac{1}{2}}_{\beta}(0)\Rightarrow \frac{(p_0+p_3)}{|p|}u^{\mathsmaller{-\frac{1}{2}}}_{\alpha} (p)v^{\mathsmaller{\frac{1}{2}}}_{\beta}(p)\,\,\,{\rm and}\,\,\,\,
u^{\frac{1}{2}}_{\alpha} (0)v^{-\frac{1}{2}}_{\beta}(0)\Rightarrow \frac{(p_0-p_3)}{|p|}u^{\mathsmaller{-\frac{1}{2}}}_{\alpha} (p)v^{\mathsmaller{\frac{1}{2}}}_{\beta}(p). \label{Vc}
 \end{eqnarray}
As we consider the product of spinors whose orientations are associated with the propagation direction of the paired particle, this inertial effect has to be taken into account. 
%where $V^b(p)$ is the vector that is observed in the moving frame of the particle and $V^a(p)$ is measured in the rest frame. 
We may rearrange the results in Eq. (\ref{produv}) to obtain the respective spinor product as follows
 \begin{eqnarray*}
u_\alpha^\mathsmaller{\frac{1}{2}}(p) v_\beta^\mathsmaller{\frac{1}{2}}(p)={\mathsmaller{\frac{1}{2}}}\gamma^0(\gamma^1+i\gamma^2)\left(\slashed{p}-m\right),&&
u_\alpha^\mathsmaller{-\frac{1}{2}}(p) v_\beta^\mathsmaller{-\frac{1}{2}}(p)=-{\mathsmaller{\frac{1}{2}}}\gamma^0(\gamma^1-i\gamma^2)\left(\slashed{p}-m\right),\\
u_\alpha^\mathsmaller{\frac{1}{2}}(p) v_\beta^\mathsmaller{-\frac{1}{2}}(p)=-{\mathsmaller{\frac{1}{2}}}(\gamma^0\gamma^5+\gamma^3)\left(\slashed{p}-m\right),&&
u_\alpha^\mathsmaller{-\frac{1}{2}}(p) v_\beta^\mathsmaller{\frac{1}{2}}(p)={\mathsmaller{\frac{1}{2}}}(\gamma^0\gamma^5-\gamma^3)\left(\slashed{p}-m\right).
 \end{eqnarray*}
Some identities between gamma matrices are provided for those in the second line of the above:
\begin{eqnarray*} 
\left.\gamma^0\gamma^5\pm\gamma^3\right.&=&\left(-\gamma^0\pm\gamma^3\right)
\mathsmaller{\frac{\left(1-\gamma^5\right)}{2}} +\left(\gamma^0\pm\gamma^3\right)
\mathsmaller{\frac{\left(1+\gamma^5\right)}{2}}.
 \end{eqnarray*} 
The above identities can be easily checked in either Dirac or chiral representation, while the latter one is adopted in this paper. It is noticed that the projection matrices, $\mathsmaller{\frac{1\pm\gamma^5}{2}}$, of the right- and left-handed fermions separate from each other.  This is an important step as we will know later that right- and left-handed fermions interact separately with different polarization states of vector bosons. After applying the above formulas on the components in the column vector in Eq. (\ref{Vb}) with polarization vectors that are defined in the appendix, we may obtain
\begin{eqnarray}
&&\hspace{-2cm}\frac{(p_0+p_3)}{|p|} u_\alpha^\mathsmaller{{-\frac{1}{2}}}(p) v_\beta^\mathsmaller{\frac{1}{2}}(p)=\label{uveq7}-\mathsmaller{\frac{1}{2}}\sum_{\lambda=0,3}\left\{\mathsmaller{(-1)}^\lambda\epsilon^{\lambda,\mu}_R\gamma_\mu\right\}\mathsmaller{\frac{\left(1-\gamma^5\right)}{2}}\left(\slashed{p}-m\right)\nonumber\\
&&\hspace{6cm}-\mathsmaller{\frac{1}{2}}\sum_{\lambda=0,3}\left\{(-1)^\lambda\epsilon^{\lambda,\mu}_L\gamma_\mu\right\}\mathsmaller{\frac{\left(1+\gamma^5\right)}{2}}\left(\slashed{p}-m\right)%|_{\alpha\beta}
,\nonumber\\ 
%\end{eqnarray}
%\begin{eqnarray}
&&\hspace{-2cm}\frac{(p_0-p_3)}{|p|} u_\alpha^\mathsmaller{\frac{1}{2}}(p) v_\beta^\mathsmaller{-\frac{1}{2}}(p)=\mathsmaller{\frac{1}{2}}\sum_{\lambda=0,3}\left\{\epsilon^{\lambda,\mu}_R\gamma_\mu\right\}\mathsmaller{\frac{\left(1-\gamma^5\right)}{2}}\left(\slashed{p}-m\right)+\mathsmaller{\frac{1}{2}}\sum_{\lambda=0,3}\left\{  \epsilon^{\lambda,\mu}_L\gamma_\mu\right\}\mathsmaller{\frac{\left(1+\gamma^5\right)}{2}}\left(\slashed{p}-m\right),
\hspace{.5cm}%|_{\alpha\beta}
\label{uveq8}
\end{eqnarray}
where $\epsilon^{\lambda,\mu}_{R,L}$ represents the R- or L-type polarization states. Here one thing that has to be emphasized is that $\epsilon^{\lambda,\mu}_{L}$ may not be physical polarization states as its longitudinal polarization vector is $\epsilon^{3,\mu}_R=(\frac{p^3}{|p|},0,0,-\frac{p^0}{|p|})$, not orthogonal to the propagation vector ${p}^\mu=(p^0,0,0,p^3)$, and it means the Ward identity could be violated. There will be
more discussions in Section \ref{ewlike} with some examples. %The trivial cases for the $x$- and $y$-dimensions are 
% \begin{eqnarray*}
%u_\alpha^\mathsmaller{\frac{1}{2}}(p) v_\beta^\mathsmaller{\frac{1}{2}}(p)=\frac{1}{2}\gamma^0(\gamma^1+i\gamma^2)\left(\slashed{p}-|p|\right),\\
%u_\alpha^\mathsmaller{-\frac{1}{2}}(p) v_\beta^\mathsmaller{-\frac{1}{2}}(p)=-\frac{1}{2}\gamma^0(\gamma^1-i\gamma^2)\left(\slashed{p}-|p|\right),
% \end{eqnarray*}

%\begin{figure}[t]
%\begin{center}
%   \subfigure[]{\includegraphics[height=4cm,width=4cm]{ATheoryOfVacuumGraphics/uvpairing.eps}}
%   \hspace*{0.0\textwidth}
%   \subfigure[]{\includegraphics[height=4cm,width=4cm]{ATheoryOfVacuumGraphics/uupairing.eps}}
%   \hspace*{0.0\textwidth}
%   \subfigure[]{\includegraphics[height=4cm,width=4cm]{ATheoryOfVacuumGraphics/vvpairing.eps}}
%\vspace{-0.0cm}
%\caption{\small  (a) The pairing of the particle and anti-particle. The spinors $u(p)$ and $v(p)$ are for fermion and anti-fermion. The panels of (b) and (c) are showing the pairing of two fermions and two anti-fermions; these two cases are ruled out as each two carries the same charges and the electrically repulsive force will not make them easy to form a pair. }
%  \label{Fig:SB}
%\end{center}
%\end{figure}
%%%%%%%%%%%%%%%%%%%%%%%%%%%%%%%%%%%%%%%%%%%%%%%%%%%%%%%%%%%%%%%%%%%%%%%%%%%%%%%%%%%
%%%%%%%%%%%%%%%%%%%%%%%%%%%%%%%%%%%%%%%%%%%%%%%%%%%%%%%%%%%%%%%%%%%%%%%%%%%%%%%%%%%
\section{Vector Bosons}
\subsection{$W^\pm$ bosons}
\subsubsection{Polarization states in $x$- and $y$-directions}
Similar to the definitions in the previous section, the propagation direction of a vector boson is assumed to be $p^\mu=(p^0,0,0,p^3)$, and each fermion shares half of the momentum, $\mathsmaller{\frac{p}{2}}$. Intuitively, the product of two fermion operator is proportional to 
 \begin{eqnarray}\label{AAdagger}\hspace{-.7cm}
\psi^r_{A,\alpha} \overline{\psi}^{s}_{A,\beta} (x) \propto\,%\frac{1}{\beta}\sum_{n={\rm even}} \int \frac{d^3 {\bf p}}{(2\pi)^3}
a^{r}_{\frac{\omega_n}{2},\frac{\bf p}{2}}c^{s}_{\frac{\omega_n}{2},\frac{\bf p}{2}} u_\alpha^r(\mathsmaller{\frac{p}{2}}) v_\beta^s(\mathsmaller{\frac{p}{2}})e^{-i\omega_n\tau+i{\bf p\cdot x}}+c^{r\dagger}_{\frac{\omega_n}{2},\frac{\bf p}{2}}a^{s,\dagger}_{\frac{\omega_n}{2},\frac{\bf p}{2}} v_\alpha^r(\mathsmaller{\frac{p}{2}}) u_\beta^r(\mathsmaller{\frac{p}{2}})e^{i\omega_n\tau-i{\bf p\cdot x}},
\end{eqnarray}
where we ignore the summation and the integral signs $\mathsmaller{\frac{1}{\beta}\sum_n \int \frac{d^3{\bf p}}{(2\pi)^3}}$. There is no such product of operators, $a^{r}_{\frac{\omega_n}{2},\frac{\bf p}{2}}c^{s\dagger}_{\frac{\omega_n}{2},\frac{\bf p}{2}}$ or $a^{r,\dagger}_{\frac{\omega_n}{2},\frac{\bf p}{2}}c^{s}_{\frac{\omega_n}{2},\frac{\bf p}{2}}$, since their exponential factors happen to cancel each other and no propagation could happen. The rest are as follows
 \begin{eqnarray*}\label{BBdagger}
\hspace{-.7cm}\psi^r_{B,\alpha} \overline{\psi}^{s}_{B,\beta} (x) \propto%\frac{1}{\beta}\sum_{n={\rm even}} \int \frac{d^3 {\bf p}}{(2\pi)^3}
b^{r}_{\frac{\omega_n}{2},\frac{\bf p}{2}}d^{s}_{\frac{\omega_n}{2},\frac{\bf p}{2}} u_\alpha^r(\mathsmaller{\frac{p}{2}}) v_\beta^s(\mathsmaller{\frac{p}{2}})e^{-i\omega_n\tau+i{\bf p\cdot x}}+d^{r\dagger}_{\frac{\omega_n}{2},\frac{\bf p}{2}}b^{s,\dagger}_{\frac{\omega_n}{2},\frac{\bf p}{2}} v_\alpha^r(\mathsmaller{\frac{p}{2}}) u_\beta^r(\mathsmaller{\frac{p}{2}})e^{i\omega_n\tau-i{\bf p\cdot x}},\\
\hspace{-.7cm}\psi^r_{A,\alpha} \overline{\psi}^{s}_{B,\beta} (x) \propto%\frac{1}{\beta}\sum_{n={\rm even}} \int \frac{d^3 {\bf p}}{(2\pi)^3}
a^{r}_{\frac{\omega_n}{2},\frac{\bf p}{2}}d^{s}_{\frac{\omega_n}{2},\frac{\bf p}{2}} u_\alpha^r(\mathsmaller{\frac{p}{2}}) v_\beta^s(\mathsmaller{\frac{p}{2}})e^{-i\omega_n\tau+i{\bf p\cdot x}}+c^{r\dagger}_{\frac{\omega_n}{2},\frac{\bf p}{2}}b^{s,\dagger}_{\frac{\omega_n}{2},\frac{\bf p}{2}} v_\alpha^r(\mathsmaller{\frac{p}{2}}) u_\beta^r(\mathsmaller{\frac{p}{2}})e^{i\omega_n\tau-i{\bf p\cdot x}},\\
\hspace{-.7cm}\psi^r_{B,\alpha} \overline{\psi}^{s}_{A,\beta} (x) \propto%\frac{1}{\beta}\sum_{n={\rm even}} \int \frac{d^3 {\bf p}}{(2\pi)^3}
b^{r}_{\frac{\omega_n}{2},\frac{\bf p}{2}}c^{s}_{\frac{\omega_n}{2},\frac{\bf p}{2}} u_\alpha^r(\mathsmaller{\frac{p}{2}}) v_\beta^s(\mathsmaller{\frac{p}{2}})e^{-i\omega_n\tau+i{\bf p\cdot x}}+d^{r\dagger}_{\frac{\omega_n}{2},\frac{\bf p}{2}}a^{s,\dagger}_{\frac{\omega_n}{2},\frac{\bf p}{2}} v_\alpha^r(\mathsmaller{\frac{p}{2}}) u_\beta^r(\mathsmaller{\frac{p}{2}})e^{i\omega_n\tau-i{\bf p\cdot x}}.
\end{eqnarray*}
An extra factor has to be taken into account, that's the propagator of a fermion %in the imaginary-time, $\frac{1}{i\omega_n-\xi_{\bf p}}$, or equivalently 
$\frac{\gamma_0}{\slashed{p}/2-m}$;  %The Fourier sum of the propagator is $\frac{1}{\beta}\sum_{\rm odd}\frac{e^{-i\omega_n \tau}}{i\omega_n-\xi_{\bf p}}=-e^{-\xi_{\bf p}\tau}\left(1-n_{\rm F}(\xi_{\bf p})\right)$, 
 it tells a fermion the possibility to find an anti-fermion to pair up with each other. As we know the two-point correlation function, $\langle \overline{\psi}(x)\psi(y)\rangle $, it represents the probability of the particle propagating from $y$ to $x$, and it is proportional to $\frac{i}{\slashed{p}/2-m}$ in the momentum space. %the fermion for an energy $\xi_{\bf p}\,(=\sqrt{{\bf p}^2+m^2})$. 
However, as we may observe from Eq. (\ref{produv}), the transverse polarization vectors in $x$- and $y$-directions have one more $\gamma^0$ than that in longitudinal $z$-direction. The additional factor, which states the pairing probability, has to be $\frac{1}{\slashed{p}/2-m}$ in the $x$- and $y$-directions to make a fermion current, $\overline{\psi}\gamma^\mu\psi$, a 4-vector and also secure the conservation of angular momentum. Now  consider the state $|1,1\rangle$ and  $|1,-1\rangle$, which are related to the $x$- and $y$-directions, and define the pairing state of the field-$A$ and the field-$B$:% with a factor, $\frac{1}{\sqrt{\beta^3 V}}$, added to make the dimension of the operators consistent:
 \begin{eqnarray}
&&\hspace{-1.5cm}\langle \langle\psi^{{\mathsmaller{\frac{1}{2}}}}_{A,\alpha} \overline{\psi}^{\,\,{{\mathsmaller{\frac{1}{2}}}}}_{B,\beta} (x)\rangle\rangle%=\langle \langle\psi^{{\mathsmaller{\frac{1}{2}}}}_{A,\alpha} \psi^{\dagger,{{\mathsmaller{\frac{1}{2}}}}}_{B,\gamma} (x)(\gamma^0)_{\gamma\beta}\rangle\rangle
\equiv
\frac{1}{\beta}\sum_{n={\rm even}} \int \frac{d^3 {\bf p}}{(2\pi)^3}\left(a^{\mathsmaller{{\mathsmaller{\frac{1}{2}}}}}_{\frac{\omega_n}{2},\frac{\bf p}{2}}d^{\mathsmaller{{\mathsmaller{\frac{1}{2}}}}}_{\frac{\omega_n}{2},\frac{\bf p}{2}} u_\alpha^{\mathsmaller{{\mathsmaller{\frac{1}{2}}}}}(\mathsmaller{\frac{p}{2}}) v_\sigma^{\mathsmaller{{\mathsmaller{\frac{1}{2}}}}}(\mathsmaller{\frac{p}{2}}) \left(\frac{1}{\mathsmaller{\frac{\slashed{p}}{2}-m}}\right)_{\sigma\gamma}(\gamma^0)_{\gamma\beta}e^{-i\omega_n\tau+i{\bf p\cdot x}}\right.\nonumber\\
&&\left.\hspace{3cm}+c^{{\mathsmaller{{\mathsmaller{\frac{1}{2}}}}}\dagger}_{\frac{\omega_n}{2},\frac{\bf p}{2}}b^{{\mathsmaller{{\mathsmaller{\frac{1}{2}}}}}\dagger}_{\frac{\omega_n}{2},\frac{\bf p}{2}} v_\sigma^{\mathsmaller{{\mathsmaller{\frac{1}{2}}}}}(\mathsmaller{\frac{p}{2}}) u_\gamma^{\mathsmaller{{\mathsmaller{\frac{1}{2}}}}}(\mathsmaller{\frac{p}{2}})\left(\frac{1}{\mathsmaller{\frac{\slashed{p}}{2}-m}}\right)_{\sigma\alpha}(\gamma^0)_{\gamma\beta}e^{i\omega_n\tau-i{\bf p\cdot x}}\right),
\label{phiA1}\nonumber\\
% \end{eqnarray}
% \begin{eqnarray}
&&\hspace{-1.5cm}\langle \langle\psi^{-{\mathsmaller{\frac{1}{2}}}}_{A,\alpha} \overline{\psi}^{{\,\,-{\mathsmaller{\frac{1}{2}}}}}_{B,\beta} (x)\rangle\rangle \equiv\frac{1}{\beta}\sum_{n={\rm even}} \int \frac{d^3 {\bf p}}{(2\pi)^3}\left(a^{-{\mathsmaller{\frac{1}{2}}}}_{\frac{\omega_n}{2},\frac{\bf p}{2}}d^{-{\mathsmaller{\frac{1}{2}}}}_{\frac{\omega_n}{2},\frac{\bf p}{2}} u_\alpha^{-{\mathsmaller{\frac{1}{2}}}}(\mathsmaller{\frac{p}{2}}) v_\beta^{-{\mathsmaller{\frac{1}{2}}}}(\mathsmaller{\frac{p}{2}}) \left(\frac{1}{\mathsmaller{\frac{\slashed{p}}{2}-m}}\right)_{\beta\gamma}(\gamma^0)_{\gamma\beta}e^{-i\omega_n\tau+i{\bf p\cdot x}}\right.\nonumber\\
&&\left.\hspace{3cm}+c^{-{\mathsmaller{\frac{1}{2}}}\dagger}_{\frac{\omega_n}{2},\frac{\bf p}{2}}b^{-{\mathsmaller{\frac{1}{2}}},\dagger}_{\frac{\omega_n}{2},\frac{\bf p}{2}} v_\beta^{-{\mathsmaller{\frac{1}{2}}}}(\mathsmaller{\frac{p}{2}}) u_\gamma^{-{\mathsmaller{\frac{1}{2}}}}(\mathsmaller{\frac{p}{2}})\left(\frac{1}{\mathsmaller{\frac{\slashed{p}}{2}-m}}\right)_{\beta\alpha}(\gamma^0)_{\gamma\beta}e^{i\omega_n\tau-i{\bf p\cdot x}}\right).\nonumber
\end{eqnarray}
Sum over the above two formulas, so that two of the transverse polarization states can be constructed:
\begin{eqnarray}
\hspace{-.5cm}\langle \langle\psi^{{\mathsmaller{\frac{1}{2}}}}_{A,\alpha} \overline{\psi}^{\,\,{{\mathsmaller{\frac{1}{2}}}}}_{B,\beta} (x)\rangle\rangle+\langle \langle\psi^{-{\mathsmaller{\frac{1}{2}}}}_{A,\alpha} \overline{\psi}^{\,\,{-{\mathsmaller{\frac{1}{2}}}}}_{B,\beta} (x)\rangle\rangle=\left(\gamma_\mu\right)_{\alpha\gamma} \sum_{\lambda=1,2}W_-^{\lambda,\mu}\mathsmaller{\frac{(1-\gamma^5)_{\gamma\beta}}{2}}+\left(\gamma_\mu\right)_{\alpha\gamma} \sum_{\lambda=1,2} W_-^{\lambda,\mu}\mathsmaller{\frac{(1+\gamma^5)_{\gamma\beta}}{2}}. \label{sumW12m}
\end{eqnarray}
The terms with $\gamma^5$-matrix actually cancel out; we will keep it in this form due to a convenience reason for the rest two polarizations, which will be derived later. The vector boson, $W_-^\mu$, is defined for two polarizations, $\lambda=1,2$, as follows
 \begin{eqnarray*}
W_-^{\lambda,\mu}&\equiv&\frac{1}{\beta}\sum_{n={\rm even}} \int \frac{d^3 {\bf p}}{(2\pi)^3}\frac{1}{\sqrt{2}}\left(  a^{\mathsmaller{W^\pm},\lambda}_{\omega_n,{\bf p}}e^{-i\omega_n\tau+i{\bf p\cdot x}}+b^{\mathsmaller{W^\pm},\lambda\dagger}_{\omega_n,{\bf p}}e^{i\omega_n\tau-i{\bf p\cdot x}}\right)\epsilon^{\lambda,\mu}, \label{W12}
 \end{eqnarray*}
where the creation and annihilation operators of $W_-^\mu$ boson are
\begin{eqnarray}\begin{matrix}
\hspace{-.7cm}a^{\mathsmaller{W^\pm},1}_{\omega_n,{\bf p}}\equiv\frac{1}{\sqrt{2}}\left(a^{{\mathsmaller{\frac{1}{2}}}}_{\frac{\omega_n}{2},\frac{\bf p}{2}}d^{{\mathsmaller{\frac{1}{2}}}}_{\frac{\omega_n}{2},\frac{\bf p}{2}}-a^{-{\mathsmaller{\frac{1}{2}}}}_{\frac{\omega_n}{2},\frac{\bf p}{2}}d^{-{\mathsmaller{\frac{1}{2}}}}_{\frac{\omega_n}{2},\frac{\bf p}{2}} \right),&
b^{\mathsmaller{W^\pm},1\dagger}_{\omega_n,{\bf p}}\equiv\frac{1}{\sqrt{2}}\left(c^{{\mathsmaller{\frac{1}{2}}}\dagger}_{\frac{\omega_n}{2}\frac{\bf p}{2}}b^{{\mathsmaller{\frac{1}{2}}},\dagger}_{\frac{\omega_n}{2},\frac{\bf p}{2}}-c^{-{\mathsmaller{\frac{1}{2}}}\dagger}_{\frac{\omega_n}{2},\frac{\bf p}{2}}b^{-{\mathsmaller{\frac{1}{2}}}\dagger}_{\frac{\omega_n}{2},\frac{\bf p}{2}} \right),\\
\hspace{-.7cm}a^{\mathsmaller{W^\pm},2}_{\omega_n,{\bf p}}\equiv\frac{i}{\sqrt{2}}\left(a^{{\mathsmaller{\frac{1}{2}}}}_{\frac{\omega_n}{2},\frac{\bf p}{2}}d^{{\mathsmaller{\frac{1}{2}}}}_{\frac{\omega_n}{2},\frac{\bf p}{2}}+a^{-{\mathsmaller{\frac{1}{2}}}}_{\frac{\omega_n}{2},\frac{\bf p}{2}}d^{-{\mathsmaller{\frac{1}{2}}}}_{\frac{\omega_n}{2},\frac{\bf p}{2}} \right),&
b^{\mathsmaller{W^\pm},2\dagger}_{\omega_n,{\bf p}}\equiv\frac{-i}{\sqrt{2}}\left(c^{{\mathsmaller{\frac{1}{2}}}\dagger}_{\frac{\omega_n}{2},\frac{\bf p}{2}}b^{{\mathsmaller{\frac{1}{2}}}\dagger}_{\frac{\omega_n}{2},\frac{\bf p}{2}}+c^{-{\mathsmaller{\frac{1}{2}}}\dagger}_{\frac{\omega_n}{2},\frac{\bf p}{2}}b^{-{\mathsmaller{\frac{1}{2}}}\dagger}_{\frac{\omega_n}{2},\frac{\bf p}{2}} \right).
\end{matrix}\label{WminusCA12}
 \end{eqnarray}
So far, it is no clear how the vector boson plays the role of a $W_-$ boson, and at this point we may simply regard it as a notation without directly relating it to real gauge bosons; this will be discussed in Section 4 as all possible interactions are arranged to form a group structure of $SU(2)_L\times U(1)_Y$.  The same step can also applies on 
 \begin{eqnarray}
&&\hspace{-1.5cm}\langle \langle\psi^{{\mathsmaller{\frac{1}{2}}}}_{B,\alpha} \overline{\psi}^{\,\,{{\mathsmaller{\frac{1}{2}}}}}_{A,\beta} (x)\rangle\rangle%=\langle \langle\psi^{{\mathsmaller{\frac{1}{2}}}}_{A,\alpha} \psi^{\dagger,{{\mathsmaller{\frac{1}{2}}}}}_{B,\gamma} (x)(\gamma^0)_{\gamma\beta}\rangle\rangle
\equiv
\frac{1}{\beta}\sum_{n={\rm even}} \int \frac{d^3 {\bf p}}{(2\pi)^3}\left(b^{\mathsmaller{{\mathsmaller{\frac{1}{2}}}}}_{\frac{\omega_n}{2},\frac{\bf p}{2}}c^{\mathsmaller{{\mathsmaller{\frac{1}{2}}}}}_{\frac{\omega_n}{2},\frac{\bf p}{2}} u_\alpha^{\mathsmaller{{\mathsmaller{\frac{1}{2}}}}}(\mathsmaller{\frac{p}{2}}) v_\sigma^{\mathsmaller{{\mathsmaller{\frac{1}{2}}}}}(\mathsmaller{\frac{p}{2}}) \left(\frac{1}{\mathsmaller{\frac{\slashed{p}}{2}-m}}\right)_{\sigma\gamma}(\gamma^0)_{\gamma\beta}e^{-i\omega_n\tau+i{\bf p\cdot x}}\right.\nonumber\\
&&\left.\hspace{3cm}+d^{{\mathsmaller{{\mathsmaller{\frac{1}{2}}}}}\dagger}_{\frac{\omega_n}{2},\frac{\bf p}{2}}a^{{\mathsmaller{{\mathsmaller{\frac{1}{2}}}}}\dagger}_{\frac{\omega_n}{2},\frac{\bf p}{2}} v_\sigma^{\mathsmaller{{\mathsmaller{\frac{1}{2}}}}}(\mathsmaller{\frac{p}{2}}) u_\gamma^{\mathsmaller{{\mathsmaller{\frac{1}{2}}}}}(\mathsmaller{\frac{p}{2}})\left(\frac{1}{\mathsmaller{\frac{\slashed{p}}{2}-m}}\right)_{\sigma\alpha}(\gamma^0)_{\gamma\beta}e^{i\omega_n\tau-i{\bf p\cdot x}}\right),
\label{phiA1}\nonumber
\end{eqnarray}
\begin{eqnarray}
&&\hspace{-1.5cm}\langle \langle\psi^{-{\mathsmaller{\frac{1}{2}}}}_{B,\alpha} \overline{\psi}^{{\,\,-{\mathsmaller{\frac{1}{2}}}}}_{A,\beta} (x)\rangle\rangle \equiv\frac{1}{\beta}\sum_{n={\rm even}} \int \frac{d^3 {\bf p}}{(2\pi)^3}\left(b^{-{\mathsmaller{\frac{1}{2}}}}_{\frac{\omega_n}{2},\frac{\bf p}{2}}c^{-{\mathsmaller{\frac{1}{2}}}}_{\frac{\omega_n}{2},\frac{\bf p}{2}} u_\alpha^{-{\mathsmaller{\frac{1}{2}}}}(\mathsmaller{\frac{p}{2}}) v_\beta^{-{\mathsmaller{\frac{1}{2}}}}(\mathsmaller{\frac{p}{2}}) \left(\frac{1}{\mathsmaller{\frac{\slashed{p}}{2}-m}}\right)_{\beta\gamma}(\gamma^0)_{\gamma\beta}e^{-i\omega_n\tau+i{\bf p\cdot x}}\right.\nonumber\\
&&\left.\hspace{3cm}+d^{\mathsmaller{-{\mathsmaller{\frac{1}{2}}}\dagger}}_{\frac{\omega_n}{2},\frac{\bf p}{2}}a^{\mathsmaller{-{\mathsmaller{\frac{1}{2}}},\dagger}}_{\frac{\omega_n}{2},\frac{\bf p}{2}} v_\beta^\mathsmaller{-{\mathsmaller{\frac{1}{2}}}}(\mathsmaller{\frac{p}{2}}) u_\gamma^\mathsmaller{-{\mathsmaller{\frac{1}{2}}}}(\mathsmaller{\frac{p}{2}})\left(\frac{1}{\mathsmaller{\frac{\slashed{p}}{2}-m}}\right)_{\beta\alpha}(\gamma^0)_{\gamma\beta}e^{i\omega_n\tau-i{\bf p\cdot x}}\right).\nonumber
\end{eqnarray}
The sum of the above formulas is defined as follows
\begin{eqnarray}
\hspace{-.7cm}\langle \langle\psi^{{\mathsmaller{\frac{1}{2}}}}_{B,\alpha} \overline{\psi}^{\,\,{{\mathsmaller{\frac{1}{2}}}}}_{A,\beta} (x)\rangle\rangle+\langle \langle\psi^{-{\mathsmaller{\frac{1}{2}}}}_{B,\alpha} \overline{\psi}^{\,\,{-{\mathsmaller{\frac{1}{2}}}}}_{A,\beta} (x)\rangle\rangle=\left(\gamma_\mu\right)_{\alpha\gamma} \sum_{\lambda=1,2}W_+^{\lambda,\mu}\mathsmaller{\frac{(1-\gamma^5)_{\gamma\beta}}{2}}+\left(\gamma_\mu\right)_{\alpha\gamma} \sum_{\lambda=1,2} W_+^{\lambda,\mu}\mathsmaller{\frac{(1+\gamma^5)_{\gamma\beta}}{2}},\,\,\label{sumW12p}
\end{eqnarray}
where, for $\lambda=1,2$,
\begin{eqnarray*}
W_+^{\lambda,\mu}&\equiv&\frac{1}{\beta}\sum_{n={\rm even}} \int \frac{d^3 {\bf p}}{(2\pi)^3}\frac{1}{\sqrt{2}}\left(  b^{\mathsmaller{W^\pm},\lambda}_{\omega_n,{\bf p}}e^{-i\omega_n\tau+i{\bf p\cdot x}}+a^{\mathsmaller{W^\pm},\lambda\dagger}_{\omega_n,{\bf p}}e^{i\omega_n\tau-i{\bf p\cdot x}}\right)\epsilon^{\lambda,\mu}.\label{W12p}
 \end{eqnarray*}
The creation and annihilation operators,  $b^{\mathsmaller{W^\pm},\lambda}_{\omega_n,{\bf p}}$ and $a^{\mathsmaller{W^\pm},\lambda\dagger}_{\omega_n,{\bf p}}$ can be obtained from Eq. (\ref{WminusCA12}) by taking the conjugation of them.

%%%%%%%%%%%%%%%%%%%%%%%%%%%%%%%%%%%%%%%%%%%%%%%%%%%%%%%%%%%%%%%%%%%%%%%%%%%%%%%%%%%
\subsubsection{Polarization states in $t$- and $z$-directions}
In this section, the longitudinal polarization will be formulated together with a non-physical polarization pointing at the time direction, as we have known from Section 2, they are mixed due to a Lorentz boost. The non-physical polarization is related to the gauge fixing term that is added to ensure the gauge invariance, and can be erased by choosing a suitable gauge. This is beyond the scope of this paper; we will keep it as the zeroth component of polarization as the discussion moves on. As we have assumed that the vector boson propagates in the direction $p^\mu_n=(p^0,0,0,p^3)$, in the time $t$- and $z$- directions, the inertial effect that is discussed in the previous section has to be included when the polarization states in these two directions are built. From Eq. (\ref{Vc}), the pairing of the two fermions can be defined as
 \begin{eqnarray}
&&\hspace{-2cm}\langle \langle\psi^{{\mathsmaller{\frac{1}{2}}}}_{A,\alpha} \overline{\psi}^{{\,\,-{\mathsmaller{\frac{1}{2}}}}}_{B,\beta} (x)\rangle\rangle%=\langle \langle\psi^{{\mathsmaller{\frac{1}{2}}}}_{A,\alpha} \psi^{\dagger,{-{\mathsmaller{\frac{1}{2}}}}}_{B,\gamma} (x)(\gamma_0)_{\gamma\beta}\rangle\rangle
\equiv\frac{1}{\beta}\sum_{n={\rm even}} \int \frac{d^3 {\bf p}}{(2\pi)^3}\left( \frac{(p_0-p_3)}{|p|}a^{{\mathsmaller{\frac{1}{2}}}}_{\frac{\omega_n}{2},\frac{\bf p}{2}}d^{-{\mathsmaller{\frac{1}{2}}}}_{\frac{\omega_n}{2},\frac{\bf p}{2}} u_\alpha^{{\mathsmaller{\frac{1}{2}}}}(\mathsmaller{\frac{p}{2}}) v_\sigma^{-{\mathsmaller{\frac{1}{2}}}}(\mathsmaller{\frac{p}{2}}) \left(\frac{1}{\mathsmaller{\frac{\slashed{p}}{2}}-\mathsmaller{m}}\gamma_0\right)_{\sigma\gamma}(\gamma_0)_{\gamma\beta}\right.\nonumber\\
&&\left.\hspace{1cm}\times e^{-i\omega_n\tau+{\bf p\cdot x}}+\frac{(p_0+p_3)}{|p|}c^{{\mathsmaller{\frac{1}{2}}}\dagger}_{\frac{\omega_n}{2},\frac{\bf p}{2}}b^{-{\mathsmaller{\frac{1}{2}}},\dagger}_{\frac{\omega_n}{2},\frac{\bf p}{2}} v_\sigma^{{\mathsmaller{\frac{1}{2}}}}(\mathsmaller{\frac{p}{2}}) u_\gamma^{-{\mathsmaller{\frac{1}{2}}}}(\mathsmaller{\frac{p}{2}})\left(\frac{1}{\mathsmaller{\frac{\slashed{p}}{2}}-\mathsmaller{m}}\gamma_0\right)_{\sigma\alpha}(\gamma_0)_{\gamma\beta}e^{i\omega_n\tau-i{\bf p\cdot x}}\right),\nonumber \label{ABpm1}\\
%\end{eqnarray}
 %\begin{eqnarray}
&&\hspace{-2cm}\langle \langle\psi^{-{\mathsmaller{\frac{1}{2}}}}_{A,\alpha} \overline{\psi}^{{\,\,{\mathsmaller{\frac{1}{2}}}}}_{B,\beta} (x)\rangle\rangle%=\langle \langle\psi^{{\mathsmaller{\frac{1}{2}}}}_{A,\alpha} \psi^{\dagger,{-{\mathsmaller{\frac{1}{2}}}}}_{B,\gamma} (x)(\gamma_0)_{\gamma\beta}\rangle\rangle
\equiv\frac{1}{\beta}\sum_{n={\rm even}} \int \frac{d^3 {\bf p}}{(2\pi)^3}\left( \frac{(p_0+p_3)}{|p|}a^{-{\mathsmaller{\frac{1}{2}}}}_{\frac{\omega_n}{2},\frac{\bf p}{2}}d^{{\mathsmaller{\frac{1}{2}}}}_{\frac{\omega_n}{2},\frac{\bf p}{2}} u_\alpha^{-{\mathsmaller{\frac{1}{2}}}}(\mathsmaller{\frac{p}{2}}) v_\sigma^{{\mathsmaller{\frac{1}{2}}}}(\mathsmaller{\frac{p}{2}}) \left(\frac{1}{\mathsmaller{\frac{\slashed{p}}{2}}-\mathsmaller{m}}\gamma_0\right)_{\sigma\gamma}(\gamma_0)_{\gamma\beta}\right.\nonumber\\
&&\left.\hspace{1cm}\times e^{-i\omega_n\tau+{\bf p\cdot x}}+\frac{(p_0-p_3)}{|p|}c^{-{\mathsmaller{\frac{1}{2}}}\dagger}_{\frac{\omega_n}{2},\frac{\bf p}{2}}b^{{\mathsmaller{\frac{1}{2}}},\dagger}_{\frac{\omega_n}{2},\frac{\bf p}{2}} v_\sigma^{-{\mathsmaller{\frac{1}{2}}}}(\mathsmaller{\frac{p}{2}}) u_\gamma^{{\mathsmaller{\frac{1}{2}}}}(\mathsmaller{\frac{p}{2}})\left(\frac{1}{\mathsmaller{\frac{\slashed{p}}{2}}-\mathsmaller{m}}\gamma_0\right)_{\sigma\alpha}(\gamma_0)_{\gamma\beta}e^{i\omega_n\tau-i{\bf p\cdot x}}\right).\nonumber \label{ABpm1}
\end{eqnarray}
The sum of the above formulas, which are related to the linear combinations of the states $|1,0\rangle$ and $|0,0\rangle$, constitutes a longitudinal and a time-like polarization states. It can be defined similar to Eq. (\ref{sumW12m}) and Eq. (\ref{sumW12p}) as
\begin{eqnarray}
\langle \langle\psi^{{\mathsmaller{\frac{1}{2}}}}_{A,\alpha} \psi^{\dagger,{-{\mathsmaller{\frac{1}{2}}}}}_{B,\gamma} (x)\rangle\rangle+\langle \langle\psi^{-{\mathsmaller{\frac{1}{2}}}}_{A,\alpha} \psi^{\dagger,{{\mathsmaller{\frac{1}{2}}}}}_{B,\gamma} (x)\rangle\rangle \hspace{-0cm}\equiv \gamma_\mu\sum_{\lambda=0,3}\left(W^{\lambda,\mu}_R\mathsmaller{\frac{1-\gamma^5}{2}}+W^{\lambda,\mu}_L\mathsmaller{\frac{1+\gamma^5}{2}}\right). \label{W03}
\end{eqnarray}
The newly defined components of the boson field $W_-^\mu$, the polarization states for $\lambda=0,3$, are
\begin{eqnarray}
W^{\lambda,\mu}_{-X}&=&\frac{1}{\beta}\sum_{n={\rm even}} \int \frac{d^3 {\bf p}}{(2\pi)^3}\frac{1}{\sqrt{2}}\left( a^{\mathsmaller{W^\pm},\lambda}_{\omega_n,{\bf p}}e^{-i\omega_n\tau-i{\bf p\cdot x}}+b^{\mathsmaller{W^\pm},\lambda\dagger}_{\omega_n,{\bf p}}e^{i\omega_n\tau-i{\bf p\cdot x}}\right)\epsilon_{X}^{\lambda,\mu},\hspace{1cm}\label{W03RL}
\end{eqnarray}
where $X=R,L$ polarization states, which are defined in appendix A. The creation and annihilation operators are
 \begin{eqnarray}
\begin{matrix}
a^{\mathsmaller{W^\pm},0}_{\omega_n,{\bf p}}=\frac{1}{\sqrt{2}}\left(a^{{\mathsmaller{\frac{1}{2}}}}_{\frac{\omega_n}{2},\frac{\bf p}{2}}d^{-{\mathsmaller{\frac{1}{2}}}}_{\frac{\omega_n}{2},\frac{\bf p}{2}} -a^{-{\mathsmaller{\frac{1}{2}}}}_{\frac{\omega_n}{2},\frac{\bf p}{2}}d^{{\mathsmaller{\frac{1}{2}}}}_{\frac{\omega_n}{2},\frac{\bf p}{2}} \right),&
a^{\mathsmaller{W^\pm},3}_{\omega_n,{\bf p}}=\frac{1}{\sqrt{2}}\left(a^{{\mathsmaller{\frac{1}{2}}}}_{\frac{\omega_n}{2},\frac{\bf p}{2}}d^{-{\mathsmaller{\frac{1}{2}}}}_{\frac{\omega_n}{2},\frac{\bf p}{2}} +a^{-{\mathsmaller{\frac{1}{2}}}}_{\frac{\omega_n}{2},\frac{\bf p}{2}}d^{{\mathsmaller{\frac{1}{2}}}}_{\frac{\omega_n}{2},\frac{\bf p}{2}} \right),\\
b^{\mathsmaller{W^\pm},0\dagger}_{\omega_n,{\bf p}}=-\frac{1}{\sqrt{2}}\left(c^{{\mathsmaller{\frac{1}{2}}}\dagger}_{\frac{\omega_n}{2},\frac{\bf p}{2}}b^{-{\mathsmaller{\frac{1}{2}}},\dagger}_{\frac{\omega_n}{2},\frac{\bf p}{2}} -c^{-{\mathsmaller{\frac{1}{2}}}\dagger}_{\frac{\omega_n}{2},\frac{\bf p}{2}}b^{{\mathsmaller{\frac{1}{2}}},\dagger}_{\frac{\omega_n}{2},\frac{\bf p}{2}} \right), &
b^{\mathsmaller{W^\pm},3\dagger}_{\omega_n,{\bf p}}=\frac{1}{\sqrt{2}}\left(c^{{\mathsmaller{\frac{1}{2}}}\dagger}_{\frac{\omega_n}{2},\frac{\bf p}{2}}b^{-{\mathsmaller{\frac{1}{2}}},\dagger}_{\frac{\omega_n}{2},\frac{\bf p}{2}} +c^{-{\mathsmaller{\frac{1}{2}}}\dagger}_{\frac{\omega_n}{2},\frac{\bf p}{2}}b^{{\mathsmaller{\frac{1}{2}}},\dagger}_{\frac{\omega_n}{2},\frac{\bf p}{2}} \right). \label{WminusCA03}
\end{matrix}
 \end{eqnarray}
There are other combinations for the states, $|1,0\rangle$ and $|0,0\rangle$ with $A$ and $B$ interchanged. We may also define 
 \begin{eqnarray}
&&\hspace{-2cm}\langle \langle\psi^{{\mathsmaller{\frac{1}{2}}}}_{B,\alpha} \overline{\psi}^{{\,\,-{\mathsmaller{\frac{1}{2}}}}}_{A,\beta} (x)\rangle\rangle%=\langle \langle\psi^{{\mathsmaller{\frac{1}{2}}}}_{A,\alpha} \psi^{\dagger,{-{\mathsmaller{\frac{1}{2}}}}}_{B,\gamma} (x)(\gamma_0)_{\gamma\beta}\rangle\rangle
\equiv\frac{1}{\beta}\sum_{n={\rm even}} \int \frac{d^3 {\bf p}}{(2\pi)^3}\left( \frac{(p_0-p_3)}{|p|}b^{{\mathsmaller{\frac{1}{2}}}}_{\frac{\omega_n}{2},\frac{\bf p}{2}}c^{-{\mathsmaller{\frac{1}{2}}}}_{\frac{\omega_n}{2},\frac{\bf p}{2}} u_\alpha^{{\mathsmaller{\frac{1}{2}}}}(\mathsmaller{\frac{p}{2}}) v_\sigma^{-{\mathsmaller{\frac{1}{2}}}}(\mathsmaller{\frac{p}{2}}) \left(\frac{1}{\mathsmaller{\frac{\slashed{p}}{2}}-\mathsmaller{m}}\gamma_0\right)_{\sigma\gamma}(\gamma_0)_{\gamma\beta}\right.\nonumber
\\
&&\left.\hspace{1cm}\times e^{-i\omega_n\tau+{\bf p\cdot x}}+\frac{(p_0+p_3)}{|p|}d^{{\mathsmaller{\frac{1}{2}}}\dagger}_{\frac{\omega_n}{2},\frac{\bf p}{2}}a^{-{\mathsmaller{\frac{1}{2}}},\dagger}_{\frac{\omega_n}{2},\frac{\bf p}{2}} v_\sigma^{{\mathsmaller{\frac{1}{2}}}}(\mathsmaller{\frac{p}{2}}) u_\gamma^{-{\mathsmaller{\frac{1}{2}}}}(\mathsmaller{\frac{p}{2}})\left(\frac{1}{\mathsmaller{\frac{\slashed{p}}{2}}-\mathsmaller{m}}\gamma_0\right)_{\sigma\alpha}(\gamma_0)_{\gamma\beta}e^{i\omega_n\tau-i{\bf p\cdot x}}\right),\nonumber \label{BApm1}
\\%\end{eqnarray}
 %\begin{eqnarray}
&&\hspace{-2cm}\langle \langle\psi^{-{\mathsmaller{\frac{1}{2}}}}_{B,\alpha} \overline{\psi}^{{\,\,{\mathsmaller{\frac{1}{2}}}}}_{A\beta} (x)\rangle\rangle%=\langle \langle\psi^{{\mathsmaller{\frac{1}{2}}}}_{A,\alpha} \psi^{\dagger,{-{\mathsmaller{\frac{1}{2}}}}}_{B,\gamma} (x)(\gamma_0)_{\gamma\beta}\rangle\rangle
\equiv\frac{1}{\beta}\sum_{n={\rm even}} \int \frac{d^3 {\bf p}}{(2\pi)^3}\left( \frac{(p_0+p_3)}{|p|}b^{-{\mathsmaller{\frac{1}{2}}}}_{\frac{\omega_n}{2},\frac{\bf p}{2}}c^{{\mathsmaller{\frac{1}{2}}}}_{\frac{\omega_n}{2},\frac{\bf p}{2}} u_\alpha^{-{\mathsmaller{\frac{1}{2}}}}(\mathsmaller{\frac{p}{2}}) v_\sigma^{{\mathsmaller{\frac{1}{2}}}}(\mathsmaller{\frac{p}{2}}) \left(\frac{1}{\mathsmaller{\frac{\slashed{p}}{2}}-\mathsmaller{m}}\gamma_0\right)_{\sigma\gamma}(\gamma_0)_{\gamma\beta}\right.\nonumber\\
&&\left.\hspace{1cm}\times e^{-i\omega_n\tau+{\bf p\cdot x}}+\frac{(p_0-p_3)}{|p|}d^{-{\mathsmaller{\frac{1}{2}}}\dagger}_{\frac{\omega_n}{2},\frac{\bf p}{2}}a^{{\mathsmaller{\frac{1}{2}}},\dagger}_{\frac{\omega_n}{2},\frac{\bf p}{2}} v_\sigma^{-{\mathsmaller{\frac{1}{2}}}}(\mathsmaller{\frac{p}{2}}) u_\gamma^{{\mathsmaller{\frac{1}{2}}}}(\mathsmaller{\frac{p}{2}})\left(\frac{1}{\mathsmaller{\frac{\slashed{p}}{2}}-\mathsmaller{m}}\gamma_0\right)_{\sigma\alpha}(\gamma_0)_{\gamma\beta}e^{i\omega_n\tau-i{\bf p\cdot x}}\right).\nonumber \label{BApm1}
\end{eqnarray}
The summation of the above gives for $\lambda=0,3$ and $X=R,L$,
\begin{eqnarray}
W^{\lambda,\mu}_{+X}&=&\frac{1}{\beta}\sum_{n={\rm even}} \int \frac{d^3 {\bf p}}{(2\pi)^3}\frac{1}{\sqrt{2}}\left( b^{\mathsmaller{W^\pm},\lambda}_{\omega_n,{\bf p}}e^{-i\omega_n\tau-i{\bf p\cdot x}}+a^{\mathsmaller{W^\pm},\lambda\dagger}_{\omega_n,{\bf p}}e^{i\omega_n\tau-i{\bf p\cdot x}}\right)\epsilon_{X}^{\lambda,\mu},\hspace{1cm},\label{W03RL}
\end{eqnarray}
where the creation and annihilation operators can be found to be consistent with the conjugates of those in Eq, (\ref{WminusCA03}). The definition of Eq. (\ref{W03RL}) can be extended to $\lambda=1,2$, as their polarization vectors of different chirality are assumed to be identical, $\epsilon^{\lambda,\mu}_R=\epsilon^{\lambda,\mu}_L=\epsilon^{\lambda,\mu}$ for these two polarizations.

%%%%%%%%%%%%%%%%%%%%%%%%%%%%%%%%%%%%%%%%%%%%%%%%%%%%%%%%%%%%%%%%%%%%%%%%%%%%%%%%%%%
\subsection{$W_3$ and $B$ bosons}
\subsubsection{Polarization states in $x$- and $y$-directions}
In the previous section, the pairing of different flavors are discussed, such as $\langle \langle\psi^{r}_{A,\alpha} \overline{\psi}^{{\,\,s}}_{B,\beta} (x)\rangle\rangle$ and $\langle \langle\psi^{r}_{B,\alpha} \overline{\psi}^{{\,\,s}}_{A,\beta} (x)\rangle\rangle$, and they are associated with the charged vector bosons $W^\mu_\pm$. Intuitively, for neutral vector bosons, it should involve the pairing of fermions of the same flavor, like $\langle \langle\psi^{r}_{A,\alpha} \overline{\psi}^{{\,\,s}}_{A,\beta} (x)\rangle\rangle$ and $\langle \langle\psi^{r}_{B,\alpha} \overline{\psi}^{{\,\,s}}_{B,\beta} (x)\rangle\rangle$.  By following the same steps for the charged bosons, they are defined for two of the polarizations as 
 \begin{eqnarray}
&&\hspace{-1.5cm}\langle \langle\psi^{{\mathsmaller{\frac{1}{2}}}}_{A,\alpha} \overline{\psi}^{{\,\,{\mathsmaller{\frac{1}{2}}}}}_{A,\beta} (x)\rangle\rangle=\frac{1}{\beta}\sum_{n={\rm even}} \int \frac{d^3 {\bf p}}{(2\pi)^3}\left(a^{{{{\mathsmaller{\frac{1}{2}}}}}}_\mathsmaller{\frac{\omega_n}{2},\frac{\bf p}{2}}c^{{\mathsmaller{{{\frac{1}{2}}}}}}_\mathsmaller{\frac{\omega_n}{2},\frac{\bf p}{2}} u_\alpha^{{\mathsmaller{{{\frac{1}{2}}}}}}(\mathsmaller{\frac{p}{2}}) v_\sigma^{{\mathsmaller{\frac{1}{2}}}}(\mathsmaller{\frac{p}{2}}) \left(\frac{1}{\mathsmaller{\frac{\slashed{p}}{2}}-\mathsmaller{m}}\right)_{\sigma\gamma}(\gamma^0)_{\gamma\beta}e^{-i\omega_n\tau+i{\bf p\cdot x}}\right.\nonumber\\
&&\left.\hspace{3cm}+c^{{\mathsmaller{\frac{1}{2}}}\dagger}_\mathsmaller{\frac{\omega_n}{2},\frac{\bf p}{2}}a^{{\mathsmaller{\frac{1}{2}}},\dagger}_\mathsmaller{\frac{\omega_n}{2},\frac{\bf p}{2}} v_\sigma^{{\mathsmaller{\frac{1}{2}}}}(\mathsmaller{\frac{p}{2}}) u_\gamma^{{\mathsmaller{\frac{1}{2}}}}(\mathsmaller{\frac{p}{2}})\left(\frac{1}{\mathsmaller{\frac{\slashed{p}}{2}}-\mathsmaller{m}}\right)_{\sigma\alpha} (\gamma^0)_{\gamma\beta}e^{i\omega_n\tau-i{\bf p\cdot x}}\right),\nonumber
 \end{eqnarray}
 \begin{eqnarray}
&&\hspace{-1.5cm}\langle \langle\psi^{{\mathsmaller{-\frac{1}{2}}}}_{A,\alpha} \overline{\psi}^{{\,\,{\mathsmaller{-\frac{1}{2}}}}}_{A,\beta} (x)\rangle\rangle=\frac{1}{\beta}\sum_{n={\rm even}} \int \frac{d^3 {\bf p}}{(2\pi)^3}\left(a^{{\mathsmaller{{\mathsmaller{-\frac{1}{2}}}}}}_\mathsmaller{\frac{\omega_n}{2},\frac{\bf p}{2}}c^{{\mathsmaller{{\mathsmaller{-\frac{1}{2}}}}}}_\mathsmaller{\frac{\omega_n}{2},\frac{\bf p}{2}} u_\alpha^{{\mathsmaller{{{-\frac{1}{2}}}}}}(\mathsmaller{\frac{p}{2}}) v_\sigma^{{\mathsmaller{-\frac{1}{2}}}}(\mathsmaller{\frac{p}{2}}) \left(\frac{1}{\mathsmaller{\frac{\slashed{p}}{2}}-\mathsmaller{m}}\right)_{\sigma\gamma}(\gamma^0)_{\gamma\beta}e^{-i\omega_n\tau+i{\bf p\cdot x}}\right.\nonumber\\
&&\left.\hspace{3cm}+c^{{\mathsmaller{-\frac{1}{2}}}\dagger}_\mathsmaller{\frac{\omega_n}{2},\frac{\bf p}{2}}a^{{\mathsmaller{-\frac{1}{2}}},\dagger}_\mathsmaller{\frac{\omega_n}{2},\frac{\bf p}{2}} v_\sigma^{{\mathsmaller{-\frac{1}{2}}}}(\mathsmaller{\frac{p}{2}}) u_\gamma^{{\mathsmaller{-\frac{1}{2}}}}(\mathsmaller{\frac{p}{2}})\left(\frac{1}{\mathsmaller{\frac{\slashed{p}}{2}}-\mathsmaller{m}}\right)_{\sigma\alpha} (\gamma^0)_{\gamma\beta}e^{i\omega_n\tau-i{\bf p\cdot x}}\right),\nonumber
\label{phiAA1}
\end{eqnarray}
The same definition applies on the pairs of $\langle \langle\psi^{r}_{B,\alpha} \overline{\psi}^{{\,\,s}}_{B,\beta} (x)\rangle\rangle$ with the replacements of $a^{r}_\mathsmaller{\frac{\omega_n}{2},\frac{\bf p}{2}}\Rightarrow b^{r}_\mathsmaller{\frac{\omega_n}{2},\frac{\bf p}{2}}$ and $c^{r}_\mathsmaller{\frac{\omega_n}{2},\frac{\bf p}{2}}\Rightarrow d^{r}_\mathsmaller{\frac{\omega_n}{2},\frac{\bf p}{2}}$. We may define the subtraction of $A$-$A$ and  $B$-$B$ pairs to form  $W_3^\mu$-like bosons, and the addition of the two to make a $B$-like boson. It will become clear how these two combinations deserve the roles of the given names in the next section.
For the $W_3^\mu$-like bosons,
 \begin{eqnarray}
&&\hspace{-1.5cm}\left(\gamma_\mu\right)_{\alpha\gamma} \sum_{\lambda=1,2} W_3^{\lambda,\mu}\mathsmaller{\frac{(1-\gamma^5)_{\gamma\beta}}{2}}+\left(\gamma_\mu\right)_{\alpha\gamma} \sum_{\lambda=1,2} W_3^{\lambda,\mu}\mathsmaller{\frac{(1+\gamma^5)_{\gamma\beta}}{2}}\equiv \sum_{r=\mathsmaller{\frac{1}{2}},{-\frac{1}{2}}}
\langle \langle\psi^{r}_{A,\alpha} \overline{\psi}^{\,\,{r}}_{A,\beta} (x)\rangle\rangle-\langle \langle\psi^{r}_{B,\alpha} \overline{\psi}^{\,\,r}_{B,\beta} (x)\rangle\rangle,\nonumber
\end{eqnarray}
where for $\lambda=1, 2$,
 \begin{eqnarray}
W^{\lambda,\mu}_{3}
&\equiv&\frac{1}{\beta}\sum_{n={\rm even}} \int \frac{d^3 {\bf p}}{(2\pi)^3}\frac{1}{\sqrt{2}}\left(  a^{\mathsmaller{W_3},\lambda}_{\omega_n,{\bf p}}e^{-i\omega_n\tau+i{\bf p\cdot x}}+a^{\mathsmaller{W_3},\lambda\dagger}_{\omega_n,{\bf p}}e^{i\omega_n\tau-i{\bf p\cdot x}}\right)\epsilon^{\lambda,\mu}.\label{W3boson}
 \end{eqnarray}
As for the addition of the two kinds of fermion pairs, the $B$-like bosons are defined as 
 \begin{eqnarray}
&&\hspace{-1.5cm}\left(\gamma_\mu\right)_{\alpha\gamma} \sum_{\lambda=1,2} B^{\lambda,\mu}\mathsmaller{\frac{(1-\gamma^5)_{\gamma\beta}}{2}}+\left(\gamma_\mu\right)_{\alpha\gamma} \sum_{\lambda=1,2} B^{\lambda,\mu}\mathsmaller{\frac{(1+\gamma^5)_{\gamma\beta}}{2}}\equiv \sum_{r=\mathsmaller{\frac{1}{2}},{-\frac{1}{2}}}
\langle \langle\psi^{r}_{A,\alpha} \overline{\psi}^{\,\,{r}}_{A,\beta} (x)\rangle\rangle+\langle \langle\psi^{r}_{B,\alpha} \overline{\psi}^{\,\,r}_{B,\beta} (x)\rangle\rangle,\nonumber
\end{eqnarray}
where for $\lambda=0 \sim 3$,
\begin{eqnarray}
B^{\lambda,\mu}_{}
&\equiv&\frac{1}{\beta}\sum_{n={\rm even}} \int \frac{d^3 {\bf p}}{(2\pi)^3}\frac{1}{\sqrt{2}}\left(  a^{\mathsmaller{B},\lambda}_{\omega_n,{\bf p}}e^{-i\omega_n\tau+i{\bf p\cdot x}}+a^{\mathsmaller{B},\lambda\dagger}_{\omega_n,{\bf p}}e^{i\omega_n\tau-i{\bf p\cdot x}}\right)\epsilon^{\lambda,\mu}.\label{Bboson}
 \end{eqnarray}
Their creation and annihilation and creation operators can be found as follows 
 \begin{eqnarray*}
a^{\mathsmaller{(W_3,B)},1}_{\omega_n,{\bf p}}&\equiv&\frac{1}{2}\left(a^{\mathsmaller{\frac{1}{2}}}_{\frac{\omega_n}{2},\frac{\bf p}{2}}c^{\mathsmaller{\frac{1}{2}}}_{\frac{\omega_n}{2},\frac{\bf p}{2}}-a^{\mathsmaller{{-\frac{1}{2}}}}_{\frac{\omega_n}{2},\frac{\bf p}{2}}c^{\mathsmaller{-\frac{1}{2}}}_{\frac{\omega_n}{2},\frac{\bf p}{2}} \mp b^{\mathsmaller{\frac{1}{2}}}_{\frac{\omega_n}{2},\frac{\bf p}{2}}d^{\mathsmaller{\frac{1}{2}}}_{\frac{\omega_n}{2},\frac{\bf p}{2}}\pm b^{\mathsmaller{-\frac{1}{2}}}_{\frac{\omega_n}{2},\frac{\bf p}{2}}d^{\mathsmaller{\mathsmaller{-\frac{1}{2}}}}_{\frac{\omega_n}{2},\frac{\bf p}{2}} \right),\\
%a^{\mathsmaller{W_3},1\dagger}_{\omega_n,{\bf p}}&\equiv&-\frac{1}{2}\left(c^{\frac{1}{2}\dagger}_{\frac{\omega_n}{2},\frac{\bf p}{2}}a^{\frac{1}{2}\dagger}_{\frac{\omega_n}{2},\frac{\bf p}{2}}-c^{-\frac{1}{2}\dagger}_{\frac{\omega_n}{2},\frac{\bf p}{2}}a^{-\frac{1}{2}\dagger}_{\frac{\omega_n}{2},\frac{\bf p}{2}} +d^{\frac{1}{2}\dagger}_{\frac{\omega_n}{2},\frac{\bf p}{2}}b^{\frac{1}{2}\dagger}_{\frac{\omega_n}{2},\frac{\bf p}{2}}-d^{-\frac{1}{2}\dagger}_{\frac{\omega_n}{2},\frac{\bf p}{2}}b^{-\frac{1}{2}\dagger}_{\frac{\omega_n}{2},\frac{\bf p}{2}}\right)\\
a^{\mathsmaller{(W_3,B)},2}_{\omega_n,{\bf p}}&\equiv&\frac{i}{2}\left(a^{\mathsmaller{\frac{1}{2}}}_{\frac{\omega_n}{2},\frac{\bf p}{2}}c^{\mathsmaller{\frac{1}{2}}}_{\frac{\omega_n}{2},\frac{\bf p}{2}}+a^{\mathsmaller{-\frac{1}{2}}}_{\frac{\omega_n}{2},\frac{\bf p}{2}}c^{\mathsmaller{-\frac{1}{2}}}_{\frac{\omega_n}{2},\frac{\bf p}{2}} \mp b^{\mathsmaller{\frac{1}{2}}}_{\frac{\omega_n}{2},\frac{\bf p}{2}}d^{\mathsmaller{\frac{1}{2}}}_{\frac{\omega_n}{2},\frac{\bf p}{2}}\mp b^{\mathsmaller{-\frac{1}{2}}}_{\frac{\omega_n}{2},\frac{\bf p}{2}}d^{\mathsmaller{-\frac{1}{2}}}_{\frac{\omega_n}{2},\frac{\bf p}{2}}\right),
%a^{\mathsmaller{W_3},2\dagger}_{\omega_n,{\bf p}}&\equiv&\frac{i}{2}\left(c^{\frac{1}{2}\dagger}_{\frac{\omega_n}{2},\frac{\bf p}{2}}a^{\frac{1}{2}\dagger}_{\frac{\omega_n}{2},\frac{\bf p}{2}}+c^{-\frac{1}{2}\dagger}_{\frac{\omega_n}{2},\frac{\bf p}{2}}a^{-\frac{1}{2}\dagger}_{\frac{\omega_n}{2},\frac{\bf p}{2}}+ d^{\frac{1}{2}\dagger}_{\frac{\omega_n}{2},\frac{\bf p}{2}}b^{\frac{1}{2}\dagger}_{\frac{\omega_n}{2},\frac{\bf p}{2}}+d^{-\frac{1}{2}\dagger}_{\frac{\omega_n}{2},\frac{\bf p}{2}}b^{-\frac{1}{2}\dagger}_{\frac{\omega_n}{2},\frac{\bf p}{2}}\right)
 \end{eqnarray*}
where $a^{\mathsmaller{W_3},\lambda}_{\omega_n,{\bf p}}$ corresponds to upper signs in the formulas, and $
a^{\mathsmaller{B},\lambda}_{\omega_n,{\bf p}}$ to the lower ones.

%%%%%%%%%%%%%%%%%%%%%%%%%%%%%%%%%%%%%%%%%%%%%%%%%%%%%%%%%%%%%%%%%%%%%%%%%%%%%%%%%%%
\subsubsection{Polarization states in $t$- and $z$-directions}
We follow the same procedures presented for $W^\pm$-like bosons for polarization states related to the propagation directions, along $t$- and $z$-axises. Define the products of fermion fields involving the states $|1,0\rangle$ and $|0,0\rangle$ as
 \begin{eqnarray}
&&\hspace{-2cm}\langle \langle\psi^{\frac{1}{2}}_{A,\alpha} \overline{\psi}^{\,\,{-\frac{1}{2}}}_{A,\beta} (x)\rangle\rangle\equiv\frac{1}{\beta}\sum_{n={\rm even}} \int \frac{d^3 {\bf p}}{(2\pi)^3}\left(\frac{(p_0-p_3)}{{|p|}}a^{\mathsmaller{\frac{1}{2}}}_{\frac{\omega_n}{2},\frac{\bf p}{2}}c^{\mathsmaller{-\frac{1}{2}}}_{\frac{\omega_n}{2},\frac{\bf p}{2}} u_\alpha^\mathsmaller{\frac{1}{2}}(\mathsmaller{\frac{p}{2}}) v_\sigma^\mathsmaller{-\frac{1}{2}}(\mathsmaller{\frac{p}{2}}) \left(\frac{1}{\mathsmaller{\frac{\slashed{p}}{2}}-\mathsmaller{m}}\gamma_0\right)_{\sigma\gamma}(\gamma_0)_{\gamma\beta}\right.\nonumber\\
&&\left.\hspace{1cm}\times e^{-i\omega_n\tau+i{\bf p\cdot x}}+\frac{(p_0+p_3)}{{|p|}}c^{\mathsmaller{\frac{1}{2}\dagger}}_{\frac{\omega_n}{2},\frac{\bf p}{2}}a^{\mathsmaller{-\frac{1}{2},\dagger}}_{\frac{\omega_n}{2},\frac{\bf p}{2}} v_\sigma^\mathsmaller{\frac{1}{2}}(\mathsmaller{\frac{p}{2}}) u_\gamma^\mathsmaller{-\frac{1}{2}}(\mathsmaller{\frac{p}{2}})\left(\frac{1}{\mathsmaller{\frac{\slashed{p}}{2}}-\mathsmaller{m}}\gamma_0\right)_{\sigma\alpha}(\gamma_0)_{\gamma\beta}e^{i\omega_n\tau-i{\bf p\cdot x}}\right),\nonumber\\
&&\hspace{-2cm}\langle \langle\psi^{-\frac{1}{2}}_{A,\alpha} \overline{\psi}^{\,\,{\frac{1}{2}}}_{A,\beta} (x)\rangle\rangle\equiv\frac{1}{\beta}\sum_{n={\rm even}} \int \frac{d^3 {\bf p}}{(2\pi)^3}\left(\frac{(p_0+p_3)}{|p|}a^{\mathsmaller{-\frac{1}{2}}}_{\frac{\omega_n}{2},\frac{\bf p}{2}}c^{\mathsmaller{\frac{1}{2}}}_{\frac{\omega_n}{2},\frac{\bf p}{2}} u_\alpha^\mathsmaller{-\frac{1}{2}}(\mathsmaller{\frac{p}{2}}) v_\sigma^\mathsmaller{\frac{1}{2}}(\mathsmaller{\frac{p}{2}}) \left(\frac{1}{\mathsmaller{\frac{\slashed{p}}{2}}-\mathsmaller{m}}\gamma_0\right)_{\sigma\gamma}(\gamma_0)_{\gamma\beta}\right.\nonumber\\
&&\left.\hspace{1cm}\times e^{-i\omega_n\tau+i{\bf p\cdot x}}+\frac{(p_0-p_3)}{|p|}c^{\mathsmaller{-\frac{1}{2}\dagger}}_{\frac{\omega_n}{2},\frac{\bf p}{2}}a^{\mathsmaller{\frac{1}{2},\dagger}}_{\frac{\omega_n}{2},\frac{\bf p}{2}} v_\sigma^\mathsmaller{-\frac{1}{2}}(\mathsmaller{\frac{p}{2}}) u_\gamma^\mathsmaller{\frac{1}{2}}(\mathsmaller{\frac{p}{2}})\left(\frac{1}{\mathsmaller{\frac{\slashed{p}}{2}}-\mathsmaller{m}}\gamma_0\right)_{\sigma\alpha}(\gamma_0)_{\gamma\beta}e^{i\omega_n\tau-i{\bf p\cdot x}}\right),\nonumber \label{AAmp1}
\end{eqnarray}
With the replacements of $a^{r}_\mathsmaller{\frac{\omega_n}{2},\frac{\bf p}{2}}\Rightarrow b^{r}_\mathsmaller{\frac{\omega_n}{2},\frac{\bf p}{2}}$ and $c^{r}_\mathsmaller{\frac{\omega_n}{2},\frac{\bf p}{2}}\Rightarrow d^{r}_\mathsmaller{\frac{\omega_n}{2},\frac{\bf p}{2}}$, the definition of the pairs of $\langle \langle\psi^{r}_{B,\alpha} \overline{\psi}^{{\,\,s}}_{B,\beta} (x)\rangle\rangle$ can be obtained. Similar definitions to the cases of the two transverse polarizations of $x$- and $y$-directions for these two polarizations, $\lambda=0,3$, to construct $W_3$-like and $B$-like bosons are 
 \begin{eqnarray}
&&\hspace{-1.5cm}  (\gamma_\mu)_{\alpha\gamma}\sum_{\lambda=0,3}\left(W^{\lambda,\mu}_{3,R}\mathsmaller{\frac{1-\gamma^5}{2}}+W^{\lambda,\mu}_{3,L}\mathsmaller{\frac{1+\gamma^5}{2}}\right)_{\gamma\beta} \equiv \sum_{r=\mathsmaller{\frac{1}{2}},{-\frac{1}{2}}}
\langle \langle\psi^{r}_{A,\alpha} \overline{\psi}^{\,\,{-r}}_{A,\beta} (x)\rangle\rangle-\langle \langle\psi^{r}_{B,\alpha} \overline{\psi}^{\,\,-r}_{B,\beta} (x)\rangle\rangle,\nonumber
\end{eqnarray}
where $W^{\lambda,\mu}_{3,X}$ for $X=R,L$, and $\lambda=0,3$, are given by 
%in Eq. (\ref{W3boson}), 
 \begin{eqnarray}
W^{\lambda,\mu}_{3,X}
&\equiv&\frac{1}{\beta}\sum_{n={\rm even}} \int \frac{d^3 {\bf p}}{(2\pi)^3}\frac{1}{\sqrt{2}}\left(  a^{\mathsmaller{W_3},\lambda}_{\omega_n,{\bf p}}e^{-i\omega_n\tau+i{\bf p\cdot x}}+a^{\mathsmaller{W_3},\lambda\dagger}_{\omega_n,{\bf p}}e^{i\omega_n\tau-i{\bf p\cdot x}}\right)\epsilon^{\lambda,\mu}_X.\nonumber\label{W3boson03}
 \end{eqnarray}
As for $B$-like bosons in these two polarizations, they are given by
 \begin{eqnarray}
&&\hspace{-1.5cm}  (\gamma_\mu)_{\alpha\gamma}\sum_{\lambda=0,3}\left(B^{\lambda,\mu}_{R}\mathsmaller{\frac{1-\gamma^5}{2}}+B^{\lambda,\mu}_{L}\mathsmaller{\frac{1+\gamma^5}{2}}\right)_{\gamma\beta} \equiv \sum_{r=\mathsmaller{\frac{1}{2},{-\frac{1}{2}}}}
\langle \langle\psi^{r}_{A,\alpha} \overline{\psi}^{\,\,{-r}}_{A,\beta} (x)\rangle\rangle+\langle \langle\psi^{r}_{B,\alpha} \overline{\psi}^{\,\,-r}_{B,\beta} (x)\rangle\rangle,\nonumber
\end{eqnarray}
where $B^{\lambda,\mu}_X$ are expressed in a similar way as follows 
 \begin{eqnarray}
B^{\lambda,\mu}_{X}
&\equiv&\frac{1}{\beta}\sum_{n={\rm even}} \int \frac{d^3 {\bf p}}{(2\pi)^3}\frac{1}{\sqrt{2}}\left(  a^{\mathsmaller{B},\lambda}_{\omega_n,{\bf p}}e^{-i\omega_n\tau+i{\bf p\cdot x}}+a^{\mathsmaller{B},\lambda\dagger}_{\omega_n,{\bf p}}e^{i\omega_n\tau-i{\bf p\cdot x}}\right)\epsilon^{\lambda,\mu}_X.\nonumber\label{W3boson03}
 \end{eqnarray}
%in Eq. (\ref{Bboson}).
%We may define the subtraction of $A$-$A$ and  $B$-$B$ pairs to form  $W_3^\mu$ bosons, and the addition of the two to make $B$ bosons. It will becomes clear how these two combinations deserve the roles of the given names in the next section.
The corresponding annihilation operators are shown in the below 
 \begin{eqnarray*}
a^{\mathsmaller{(W_3,B)},0}_{\omega_n,{\bf p}}&\equiv&\frac{1}{2}\left(a^{\mathsmaller{\frac{1}{2}}}_{\frac{\omega_n}{2},\frac{\bf p}{2}}c^{-\frac{1}{2}}_{\frac{\omega_n}{2},\frac{\bf p}{2}} -a^{-\frac{1}{2}}_{\frac{\omega_n}{2},\frac{\bf p}{2}}c^{\frac{1}{2}}_{\frac{\omega_n}{2},\frac{\bf p}{2}}\mp b^{\frac{1}{2}}_{\frac{\omega_n}{2},\frac{\bf p}{2}}d^{-\frac{1}{2}}_{\frac{\omega_n}{2},\frac{\bf p}{2}} \pm b^{-\frac{1}{2}}_{\frac{\omega_n}{2},\frac{\bf p}{2}}d^{\frac{1}{2}}_{\frac{\omega_n}{2},\frac{\bf p}{2}}\right),\\
a^{\mathsmaller{(W_3,B)},3}_{\omega_n,{\bf p}}&\equiv&\frac{1}{2}\left(a^{\frac{1}{2}}_{\frac{\omega_n}{2},\frac{\bf p}{2}}c^{-\frac{1}{2}}_{\frac{\omega_n}{2},\frac{\bf p}{2}} +a^{-\frac{1}{2}}_{\frac{\omega_n}{2},\frac{\bf p}{2}}c^{\frac{1}{2}}_{\frac{\omega_n}{2},\frac{\bf p}{2}}\mp b^{\frac{1}{2}}_{\frac{\omega_n}{2},\frac{\bf p}{2}}d^{-\frac{1}{2}}_{\frac{\omega_n}{2},\frac{\bf p}{2}} \mp b^{-\frac{1}{2}}_{\frac{\omega_n}{2},\frac{\bf p}{2}}d^{\frac{1}{2}}_{\frac{\omega_n}{2},\frac{\bf p}{2}} \right),
 \end{eqnarray*}
and the creation operators can be obtained by taking the conjugation of the respective annihilation operators.
%%%%%%%%%%%%%%%%%%%%%%%%%%%%%%%%%%%%%%%%%%%%%%%%%%%%%%%%%%%%%%%%%%%%%%%%%%%%%%%%%%%
%%%%%%%%%%%%%%%%%%%%%%%%%%%%%%%%%%%%%%%%%%%%%%%%%%%%%%%%%%%%%%%%%%%%%%%%%%%%%%%%%%%
\section{An Electroweak-like Theory of $SU(2)_L\times U(1)_Y$}
\label{ewlike}
In Section 2, four kinds of four-fermion interaction are presented; as we consider the possible pairing of a fermion and an anti-fermion, there are more than four types that have to be included. Fortunately, an identity is known, $\bar{u}^r_\alpha (p) v^s_\alpha (p)=\bar{v}^s_\alpha (p) u^r_\alpha (p)=0$, so that such pairing, $\langle \langle\psi^{r}_{A,\alpha} \overline{\psi}^{\,s}_{A,\alpha} \rangle\rangle$, etc., can be ignored. In general, the interactions of (a) and (b) in Eq. (\ref{4fermion}), while attached with coupling constants, can be cast into a $2\times 2$ matrix contracted by a row and a column vector 
 \begin{eqnarray}
\sum_{r,s,r',s'} 
\left(\begin{matrix}
\overline{\psi}^{\,r'}_{B,\alpha}&\overline{\psi}^{\,r'}_{A,\alpha}
\end{matrix}\right)
\left(\begin{matrix}
g_b\langle \langle\psi^{r}_{A,\alpha} \overline{\psi}^{\,s}_{A,\beta} \rangle\rangle & 
g_{c}\langle \langle\psi^{r}_{B,\alpha} \overline{\psi}^{\,s}_{A,\beta} \rangle\rangle\\
g_c \langle \langle\psi^{r}_{A,\alpha} \overline{\psi}^{\,s}_{B,\beta} \rangle\rangle& 
g_b\langle \langle\psi^{r}_{B,\alpha} \overline{\psi}^{\,s}_{B,\beta} \rangle\rangle
\end{matrix}\right)
\left(\begin{matrix}
{\psi}^{\,s'}_{B,\beta}\\
{\psi}^{\,s'}_{A,\beta}
\end{matrix}\right).\label{4ABfermion}
 \end{eqnarray} 
Two $g_b$ are assumed to be equivalent under the assumption of the symmetry between $A$- and $B$-types of particles, as well as two $g_c$. For the interactions of four the same type fermions, (c) and (d), they can also be depicted by a matrix multiplication shown in below
 \begin{eqnarray}
g_a\sum_{r,s,r',s'} 
\left(\begin{matrix}
\overline{\psi}^{\,r'}_{B,\alpha}&\overline{\psi}^{\,r'}_{A,\alpha}
\end{matrix}\right)
\left(\begin{matrix}
\langle \langle\psi^{r}_{B,\alpha} \overline{\psi}^{\,s}_{B,\beta} \rangle\rangle & 0\\
0& \langle \langle\psi^{r}_{A,\alpha} \overline{\psi}^{\,s}_{A,\beta} \rangle\rangle
\end{matrix}\right)
\left(\begin{matrix}
{\psi}^{\,s'}_{B,\beta}\\
{\psi}^{\,s'}_{A,\beta}
\end{matrix}\right).\label{4AAorBBfermions}
 \end{eqnarray}
The two coupling constants of above are made the same, as two flavors of fermions are treated equally.
The sum of the above two $2\times 2$ matrices is 
 \begin{eqnarray*}
&&\sum_{r,s} 
\left(\begin{matrix}
g_b \langle \langle\psi^{r}_{A,\alpha} \overline{\psi}^{\,s}_{A,\beta} \rangle\rangle +g_a \langle \langle\psi^{r}_{B,\alpha} \overline{\psi}^{\,s}_{B,\beta} \rangle\rangle & 
g_c \langle \langle\psi^{r}_{B,\alpha} \overline{\psi}^{\,s}_{A,\beta} \rangle\rangle\\
g_c \langle \langle\psi^{r}_{A,\alpha} \overline{\psi}^{\,s}_{B,\beta} \rangle\rangle & 
g_b \langle \langle\psi^{r}_{B,\alpha} \overline{\psi}^{\,s}_{B,\beta} \rangle\rangle
+ g_a \langle \langle\psi^{r}_{A,\alpha} \overline{\psi}^{\,s}_{A,\beta} \rangle\rangle
\end{matrix}\right),
 \end{eqnarray*}
 \begin{eqnarray*}
&=&\sum_{r,s} 
\left(\begin{matrix}
\frac{g_b-g_a}{2} \left(\langle \langle\psi^{r}_{A,\alpha} \overline{\psi}^{\,s}_{A,\beta} \rangle\rangle - \langle \langle\psi^{r}_{B,\alpha} \overline{\psi}^{\,s}_{B,\beta} \rangle\rangle\right) & 
g_c \langle \langle\psi^{r}_{B,\alpha} \overline{\psi}^{\,s}_{A,\beta} \rangle\rangle\\
g_c \langle \langle\psi^{r}_{A,\alpha} \overline{\psi}^{\,s}_{B,\beta} \rangle\rangle & 
-\frac{g_b-g_a}{2} \left( \langle \langle\psi^{r}_{A,\alpha} \overline{\psi}^{\,s}_{A,\beta} \rangle\rangle
- \langle \langle\psi^{r}_{B,\alpha} \overline{\psi}^{\,s}_{B,\beta} \rangle\rangle\right)
\end{matrix}\right)\\
&&+\frac{(g_a+g_b)}{2}\sum_{r,s} 
\left(\langle \langle\psi^{r}_{A,\alpha} \overline{\psi}^{\,s}_{A,\beta} \rangle\rangle + \langle \langle\psi^{r}_{B,\alpha} \overline{\psi}^{\,s}_{B,\beta} \rangle\rangle\right) 
\left(\begin{matrix}
1&0\\
0&1\end{matrix}\right).
 \end{eqnarray*}
It looks that the above matrix contains group structures of $SU(2)$ and $U(1)$, similar to the electroweak theory; however it has three couplings, instead of two. Once they are related to the coupling constants, $g_1$ and $g_2$, in the GSW theory, we may assign each component in the matrix a proper name of vector boson, and that is what we did in the previous sections. To summarize and combine all of the polarizations for $W^\pm$, $W_3$ and $B$ bosons, from $\lambda=0$ to 3, we obtain
 \begin{eqnarray*}
&&\sum_{r,s}\langle \langle\psi^{r}_{A,\alpha} \overline{\psi}^{\,\,s}_{B,\beta} \rangle\rangle=\left(\gamma_\mu\right)_{\alpha\gamma}\left(W^{\mu}_{-,R}\mathsmaller{\frac{1-\gamma^5}{2}}+W^{\mu}_{-,L}\mathsmaller{\frac{1+\gamma^5}{2}}\right)_{\gamma\beta}, \nonumber\\
&&\sum_{r,s}\langle \langle\psi^{r}_{B,\alpha} \overline{\psi}^{\,\,s}_{A,\beta} \rangle\rangle=\left(\gamma_\mu\right)_{\alpha\gamma}\left(W^{\mu}_{+,R}\mathsmaller{\frac{1-\gamma^5}{2}}+W^{\mu}_{+,L}\mathsmaller{\frac{1+\gamma^5}{2}}\right)_{\gamma\beta}, \nonumber\\
&&\hspace{0cm}\sum_{r,s}\left(\langle \langle\psi^{r}_{A,\alpha} \overline{\psi}^{\,\,s}_{A,\beta} \rangle\rangle-\langle \langle\psi^{r}_{B,\alpha} \overline{\psi}^{\,\,s}_{B,\beta} \rangle\rangle\right)=\sqrt{2}\left(\gamma_\mu\right)_{\alpha\gamma}\left(W^{\mu}_{3,R}\mathsmaller{\frac{1-\gamma^5}{2}}+W^{\mu}_{3,L}\mathsmaller{\frac{1+\gamma^5}{2}}\right)_{\gamma\beta}, \\
&&\hspace{0cm}\sum_{r,s}\left(\langle \langle\psi^{r}_{A,\alpha} \overline{\psi}^{\,\,s}_{A,\beta} \rangle\rangle+\langle \langle\psi^{r}_{B,\alpha} \overline{\psi}^{\,\,s}_{B,\beta} \rangle\rangle\right)=\sqrt{2}\left(\gamma_\mu\right)_{\alpha\gamma}\left(B^{\mu}_{R}\mathsmaller{\frac{1-\gamma^5}{2}}+B^{\mu}_{L}\mathsmaller{\frac{1+\gamma^5}{2}}\right)_{\gamma\beta}.
\end{eqnarray*}
We may relate the new coupling constants to those in the Standard Model, $g_1$ and $g_2$,
 \begin{eqnarray*}
\frac{g_2}{\sqrt{2}}= g_c=\mathsmaller{{g_b-g_a}},\,\, {\rm and}\, \, \,\,\frac{g_1}{\sqrt{2}} \,%Y
= \mathsmaller{{g_a+g_b}}.
 \end{eqnarray*} 
By doing this, the sum of all four fermion interactions in Eq. (\ref{4ABfermion}) and (\ref{4AAorBBfermions}) forms group structures of SU(2) and U(1). The $2\times 2$ matrix  now becomes
 \begin{eqnarray}
&=&g_2(\gamma_\mu)_{\alpha\gamma}
\left(\begin{matrix}
{\frac{1}{{2}}} \left(W^{\mu}_{3,R}\frac{1-\gamma^5}{2}+W^{\mu}_{3,L}\frac{1+\gamma^5}{2}\right)_{\gamma\beta} & 
 {\frac{1}{\sqrt{2}}}  \left(W^{\mu}_{+,R}\frac{1-\gamma^5}{2}+W^{\mu}_{+,L}\frac{1+\gamma^5}{2}\right)_{\gamma\beta} \\
{\frac{1}{\sqrt{2}}} \left(W^{\mu}_{-,R}\frac{1-\gamma^5}{2}+W^{\mu}_{-,L}\frac{1+\gamma^5}{2}\right)_{\gamma\beta}  & 
-{\frac{1}{2}} \left(W^{\mu}_{3,R}\frac{1-\gamma^5}{2}+W^{\mu}_{3,L}\frac{1+\gamma^5}{2}\right)_{\gamma\beta} 
\end{matrix}\right)\nonumber\\
&&+g_1 %Y
\sum_{\lambda=0}^{\mathsmaller 3}\left(B^{\lambda,\mu}_{R}\mathsmaller{\frac{1-\gamma^5}{2}}+B^{\lambda,\mu}_{L}\mathsmaller{\frac{1+\gamma^5}{2}}\right)_{\gamma\beta} 
\left(\begin{matrix}
\mathsmaller{\frac{1}{{2}}}&0\\
0&\mathsmaller{\frac{1}{{2}}}\end{matrix}\right),\nonumber\\
&=&\gamma_\mu \mathsmaller{\frac{1-\gamma^5}{2}}\left\{ g_2(\mathsmaller{\frac{1}{\sqrt{2}}}W^{\mu}_{+,R} T^+ +\mathsmaller{\frac{1}{\sqrt{2}}}W^{\mu}_{-,R}T^- + W^{\mu}_{3,R} T^3)+g_1 %Y
 B^{\mu}_R \right\}\nonumber\\
&&\hspace{2cm}+\gamma_\mu \mathsmaller{\frac{1+\gamma^5}{2}}\left\{ g_2( \mathsmaller{\frac{1}{\sqrt{2}}}W^{\mu}_{+,L} T^+ +\mathsmaller{\frac{1}{\sqrt{2}}}W^{\mu}_{-,L}T^- + W^{\mu}_{3,L} T^3)+g_1 %Y 
B^{\mu}_L \right\},\label{ewmatrix}
 \end{eqnarray}
where $T^+=\mathsmaller{\left(\begin{matrix} \mathsmaller{0}&\mathsmaller{1}\\ \mathsmaller{0}&\mathsmaller{0}\end{matrix}\right)}$,  $T^-=\mathsmaller{\left(\begin{matrix} \mathsmaller{0}&\mathsmaller{0}\\ \mathsmaller{1}&\mathsmaller{0}\end{matrix}\right)}$ and  $T^3=\mathsmaller{\left(\begin{matrix} \mathsmaller{\frac{1}{{2}}}&\mathsmaller{0}\\ \mathsmaller{0}&\mathsmaller{-\frac{1}{{2}}}\end{matrix}\right)}$. It is observed that the chiral symmetry of fermions is broken. Two group structures, $SU(2)_L\times U(1)_Y$ and $SU(2)_R\times U(1)_Y$,  are formed, and it shows that the left-handed fermions interact with R-type bosons and vise versa. If the vacuum is not broken, all of the bosons are massless and have only transverse polarizations. The chiral symmetry is hold for both handedness fermions, so are the Ward identity, $p\cdot \epsilon^{1,2}=0$. However, once the vacuum is broken,  the gauge invariance requires all of the three polarizations satisfying: $p\cdot \epsilon^i=0$, (for $i=1,2,3$); only $\epsilon^\mu_R$ does, and $\epsilon^\mu_L$ does not. To be specific, in our initial assumption of propagation vector $p^\mu=(p^0,0,0,p^3)$, and from the longitudinal part of the L-type polarization vectors, $p\cdot\epsilon^3_L\neq 0$. Therefore, in the second line, the group of $SU(2)_R$, where right-handed fermions couples to non-physical $W^i_L$-like ($i=1,2,3$) bosons, violates the Ward identity, so right-handed fermions appear being prohibited to interact with L-type vector bosons. In fact, there are exceptions when either $p^0$ or $p^3$ is zero for these L-type bosons to participate in the scattering with the right-handed fermions. We may notice before going further that the both sets of the polarization vectors are 4-vectors, therefore the scalar product of any two polarization vectors are scalars.  In the Møller or Bhabha scattering \cite{moller,bhabha}, the neutral massless bosons interact freely with both handedness fermions, as well as the neutral massive  bosons due to $p^0=0$ for $t$- or $u$-channel or $p^3=0$ for $s$-channel. In the $\beta$- or muon-decay \cite{betadecay,muondecay}, $p_0$ can not vanish, since the produced lepton and anti-lepton pair are physical particles. Therefore, the right-handed fermions do not couple to the L-type $W^\pm$ bosons. Moreover, if $p^3=0$, the Ward identity, $p\cdot \epsilon^3_L=0$, can also hold; in the $Z$ resonance \cite{Zresonance}, which is a $s$-channel process of $e^+ e^-\rightarrow {\rm f \bar{f}}$, a L-type $Z$- or $\gamma$-like boson can only interact with right-handed fermions through this way. %As for neutrinos, it is well known that they oscillates among three flavors \cite{neutrinooscillation}. This destroys the possibility of right-handed leptons and left-handed anti-leptons collide to create $W^\pm$ bosons in this channel, since electrons do not oscillate. %As we know, $W^\pm$ can not participate 
Based from the above discussions the parts in the $2\times 2$ matrix from $W^i_L$-like (i=1,2,3) bosons of $SU(2)_R$ can be  removed from the possible physical states of four vector bosons. Therefore in the following, for the right-handed fermions, the components of $Z_L$ and $A_L$ bosons from $W^3_L$ can be neglected.  An interesting case is the pion decay \cite{piondecay}, $\pi^-\rightarrow \mu^- \bar{\nu}_\mu$. It is also a $s$-channel. Since the observed pions have nonzero 3-momenta, $p^3\neq 0$, as well as $p^0\neq 0$ for a $s$-channel, the L-type $W^-$ bosons  can not mediate the electroweak forces and only left-handed fermions and right-handed anti-fermions can be produced.  Since photons are massless, they interact with both chiralities of fermions equally no matter the vacuum is broken or not.  As we can only measure the coupling constants of the broken phase in laboratory, we may redefine the coupling constants, $g_1$ and $g_2$, in Eq. (\ref{ewmatrix}) as $g_{1,L}$, $g_{2,L}$, $g_{1,R}$ and $g_{2,R}$ for the respective left- and right-handed fermions.
To conclude, we may remove the unphysical parts in the $2\times 2$ matrix, the L-type $W$ bosons by setting $g_{2,R}=0$, and obtain 
 \begin{eqnarray*}
=\gamma_\mu \mathsmaller{\frac{1-\gamma^5}{2}}\left( g_{2,L}(\mathsmaller{\frac{1}{\sqrt{2}}}W^{\mu}_{+,R} T^+ +\mathsmaller{\frac{1}{\sqrt{2}}}W^{\mu}_{-,R}T^- + W^{\mu}_{3,R} T^3)+g_{1,L} %Y 
B^{\mu}_R \right)
+\gamma_\mu \mathsmaller{\frac{1+\gamma^5}{2}}\left. g_{1,R} %Y
 B^{\mu}_L \right.,
 \end{eqnarray*}
where $g_{1,X}=g'Y$ for $X=L$, $Y=-\mathsmaller{\frac{1}{2}}$ and $X=R$, $Y=-1$ in order to satisfy the observation results. The factor $Y$ is the so-called weak hypercharge. Thus they correspond to the conventional notations $g$ and $g'$ as in Peskin's \cite{peskin95} with $g_{2,L}=g$. They can be expressed as below
 \begin{eqnarray*}
g_1\xRightarrow{S.B.}g'\,Y=
\begin{cases}\hspace{.2cm}
g_{1,L}=-\mathsmaller{\frac{1}{2}}\,g'\\ \hspace{.2cm}
g_{1,R}=-g'
 \end{cases}, {\rm and} \hspace{,5cm}
g_2\xRightarrow{S.B.}
\begin{cases}\hspace{.2cm}
g_{2,L}=g\\ \hspace{.2cm}
g_{2,R}=0
 \end{cases},
 \end{eqnarray*}
where S.B. stands for symmetry breaking. We may use the conventional variable change of $W^3_X$ and $B_X$ ($X=R,L$) bosons for their mass eigenstates after  symmetry breaking,
 \begin{eqnarray*}
\begin{cases}
Z^\mu_X=\mathsmaller{\frac{g}{\sqrt{{g'}^2+g^2}}}W^\mu_{3,X}-\mathsmaller{\frac{g'}{\sqrt{g'^2+g^2}}}B^\mu_{X}\\
A^\mu_X=\mathsmaller{\frac{g'}{\sqrt{g'^2+g^2}}}W^\mu_{3,X}+\mathsmaller{\frac{g}{\sqrt{g'^2+g^2}}}B^\mu_{X}
\end{cases}\hspace{.3cm}{\rm ,\,\,or \,\,}\hspace{.3cm}
 \begin{cases}
W^\mu_{3,R}=\mathsmaller{\frac{g}{\sqrt{g'^2+g^2}}}Z^\mu_{R}+\mathsmaller{\frac{g'}{\sqrt{g'^2+g^2}}}A^\mu_{R}\\
B^\mu_X\,\,\,=-\mathsmaller{\frac{g'}{\sqrt{g'^2+g^2}}}Z^\mu_{X}+\mathsmaller{\frac{g}{\sqrt{g'^2+g^2}}}A^\mu_{X},
 \end{cases}
 \end{eqnarray*}
where $W^\mu_{3,L}$ has to be  set to zero. By using the conventional definitions of these coupling constants, $e=\mathsmaller{\frac{g'g}{\sqrt{g'^2+g^2}}}$, $\cos \theta_W=\mathsmaller{\frac{g}{\sqrt{g'^2+g^2}}}$, $\sin\theta_W=\mathsmaller{\frac{g'}{\sqrt{g'^2+g^2}}}$ and the charge $Q=T^3+Y$, the $2\times 2$ matrix becomes
 \begin{eqnarray*}
&&\hspace{-1cm}=\gamma_\mu \mathsmaller{\frac{1-\gamma^5}{2}}\left\{ \mathsmaller{\frac{e}{\sin\theta_W}}(W^{\mu}_{+,R} T^+ +W^{\mu}_{-,R}T^-) +\mathsmaller{\frac{e}{\sin\theta_W\cos\theta_W}} \left(  T^3-\sin^2\theta_W Q\right)Z^\mu_{R} 
+eQA^\mu_{R}  \right\}\\
%&&\hspace{-0cm}+\gamma_\mu \mathsmaller{\frac{1+\gamma^5}{2}}\left(\mathsmaller{\frac{e}{\sin\theta_W}}(W^{\mu}_{+,L} T^+ +W^{\mu}_{-,L}T^- )+  \mathsmaller{\frac{e}{\sin\theta_W\cos\theta_W}} \left(  T^3-\sin^2\theta_W Q\right)Z^\mu_{L} +eQA^\mu_{L} \right).
&&\hspace{3cm}+\gamma_\mu \mathsmaller{\frac{1+\gamma^5}{2}}\left\{\mathsmaller{-\frac{e \sin^2\theta_W}{\sin\theta_W\cos\theta_W}} \left. Q\right.Z^\mu_{L} 
+eQA^\mu_{L} \right\}.
 \end{eqnarray*}
Only the components of $Z^\mu_L$ and $A^\mu_L$ in $B_L$-like bosons for right-handed fermions may survive from the Ward identity after symmetry breaking. This is similar to the picture that the Standard Model has given us.% except $Z_L$ only exists in the $s$-channel.
%In the first line, a group structure, similar to $SU(2)_L$, with chirality is formed, and it shows that the left-handed fermions interact with right-handed bosons due to the fact that the gauge invariance requires that $p\cdot \epsilon^i=0$, (for $i=1,2,3$). Only $\epsilon^\mu_R$ satisfies, and $\epsilon^\mu_L$ does not. To be specific, in our initial assumption of propagation vector $p^\mu=(p^0,0,0,p^3)$, and from the longitudinal part of the left-handed polarization vectors, $p\cdot\epsilon^3_L\neq 0$. On the other hand, in the second line a group of $SU(2)_R$ appears, and right-handed fermions couples to non-physical $W^\pm_L$, $Z_L$ bosons. It violates the gauge invariance, so right-handed fermions are prohibited to interact with left-handed vector bosons.  However, if $p_3=0$, the above constraint is secured again since $\epsilon^3_L$ and $p^\mu$ are orthogonal, then the gauge invariance holds. This condition happens in the $s$-channel of the collision, ${\rm e\,\bar{e}}\rightarrow {\rm f \,\bar{f}}$, so that $Z$ resonance occurs without chiral difference. As for a physical photon, it has only two transverse polarization states, $\epsilon^1$ and $\epsilon^2$, so $A^\mu_L=A^\mu_R$. Therefore what vector boson can interact with the right-handed fermions is the photon, as expected.
%%%%%%%%%%%%%%%%%%%%%%%%%%%%%%%%%%%%%%%%%%%%%%%%%%%%%%%%%%%%%%%%%%%%%%%%%%%%%%%%%%%
%%%%%%%%%%%%%%%%%%%%%%%%%%%%%%%%%%%%%%%%%%%%%%%%%%%%%%%%%%%%%%%%%%%%%%%%%%%%%%%%%%%
\section{Conclusion}
A toy model that features an electroweak-like group structure is formulated from four-fermion interactions, and it is inspired from the BCS theory whose ground state is broken due to collective pairs of electrons. The so-called Cooper pairs interact with each other and are responsible for the phenomena of superconductivity in solids.  Accordingly, pairs of a fermion and an anti-fermion are made here to form four vector bosons, and each of them behaves like $W^\pm$, $Z$ and $\gamma$ bosons, in a group structure of $SU(2)_L\times U(1)_Y$. The reason is that the theory automatically separate left- and right-handed fermions to interact respectively with vector bosons of two distinctive kinds of polarization states derived from the products of Dirac spinors. If this theory is appreciable, it raises a conceptual question about the definition of particles. It appears that a fermion and a anti-fermion make a vector boson, however, it is not quite the truth since a pair, such as $\langle \langle \psi_{A,\alpha} \overline{\psi}_{B,\beta}\rangle \rangle$, still contains a gamma matrix. It is this gamma matrix appearing in  fermion's four-current, $\overline{\psi}\gamma^\mu\psi$. The field operator of a vector boson, which is extracted from the four-fermion interaction, only possesses the oscillators, $a_{\omega_n,{\bf p}}$,
$a^\dagger_{\omega_n,{\bf p}}$, etc., and leaves its Dirac structure in the interaction vertex. This is saying 
that a vector boson is born from the condensate of a fermion and an anti-fermion pair and only inherits part of the ingredients from the pair, $\langle \langle \psi_{X,\alpha} \overline{\psi}_{Y,\beta}\rangle \rangle$. As far as we may know so far from its mathematical structure, apart from other irrelevant factors, a vector boson only carries the pair of fermion oscillators without other identities of the individual fermion, such as mass, etc., and indeed fermion pairs of creation and annihilation operators behave like bosonic ones.  In Section 2, the masses of two types of fermions,  $A$ and $B$, are suggested to be neglected while the momentum are extremely larger than their masses.  From the above point of view, a pair of fermions may be regarded as massless before a vector boson gains its mass through the Higgs mechanism.  From the field theory,  a single fermion obtains its mass from a Yukawa potential with Higgs particles. A fermion in a pair and an isolated one  are in different states, and they receive different masses through the same broken vacuum. Thus, at the moment a pair of spinors is formed, two single fermion oscillators are supposed to give up their original masses, as well as the spinor structures, and obtain a new mass together as being a vector boson until next annihilation. Nevertheless, up to this point this is an initial attempt to build up an electroweak-like theory in a condensed matter approach, and more investigations and inspirations on this subject to see if it is really applicable on real physical systems are necessary in the future.

%%%%%%%%%%%%%%%%%%%%%%%%%%%%%%%%%%%%%%%%%%%%%%%%%%%%%%%%%%%%%%%%%%%%%%%%%%%%%%%%%%%
%%%%%%%%%%%%%%%%%%%%%%%%%%%%%%%%%%%%%%%%%%%%%%%%%%%%%%%%%%%%%%%%%%%%%%%%%%%%%%%%%%%
\appendix
\section{Sums of Polarization Vectors}
The polarization vectors that are used in this paper are explained in this section. The difference from the usual ones is that two sets of polarization states for a vector boson are defined, one is for a L-type system and the other is the R-type. For a particle propagating in the direction $u^\mu=\frac{p^\mu}{|p|}=(\frac{p^0}{|p|},0,0,\frac{p^3}{|p|})$, where $|p|=\sqrt{p^2}$, the R-type set of vectors are 
\begin{eqnarray*}
\epsilon^{0,\mu}_R=(\frac{p^0}{|p|},0,0,\frac{p^3}{|p|}),&& \epsilon^{3,\mu}_R=(\frac{p^3}{|p|},0,0,\frac{p^0}{|p|}), \\
\epsilon^{1,\mu}_R=(0,1,0,0)\hspace{0.6cm},&&\epsilon^{2,\mu}_R=(0,0,1,0),
\end{eqnarray*}
and all vectors are orthogonal to each other. The first 4-vector, $\epsilon^{0,\mu}_R$, is a time-like vector. The second one, $\epsilon^{3,\mu}_R$, is the longitudinal polarization, while $\epsilon^{1,\mu}_R$ and $\epsilon^{2,\mu}_R$ are the transverse parts. All of the three polarization states are orthogonal to the 4-momentum, $(p\cdot \epsilon^i_R)=0$, for $i=1,2,3$. % If the particle is at rest, $\epsilon^3_R=(0,0,0,1)$. Then the spatial components of the three polarization vectors, by given $\epsilon^i_R=(0,\vec{v}_i)$, make a right-handed coordinate system, that's $\vec{v}_i\times \vec{v}_j=\epsilon^{ijk}\vec{v}_k$, where $\epsilon^{ijk}$ is a Levi-Civita symbol. 
As for the L-type set, it also forms a orthonormal set as follows
\begin{eqnarray*}
\epsilon^{0,\mu}_L=(-\frac{p^0}{|p|},0,0,\frac{p^3}{|p|}),&& \epsilon^{3,\mu}_L=(-\frac{p^3}{|p|},0,0,\frac{p^0}{|p|}), \\
\epsilon^{1,\mu}_L=(0,1,0,0)\hspace{0.9cm},&&\epsilon^{2,\mu}_L=(0,0,1,0).
\end{eqnarray*}
%In particle's rest frame, the longitudinal vector is $\epsilon^3_L=(0,0,0,1)$; the spatial parts of the three polarization vectors, by setting $\epsilon^i_L=(0,\vec{w}_i)$, are building a left-handed coordinate system, $\vec{w_i}\times \vec{w_j}=-\epsilon^{ijk}\vec{w}_k$. This is the reason that each set of polarizations deserves its chiral nomenclature.
An important difference from the first set is that the scalar product of $(p\cdot \epsilon^3_L)\neq0$, and it leads to the fact that the L-type polarization vectors violate the gauge invariance. Therefore, the massive left-handed vector bosons are not physical. For the massless bosons, since they just have two transverse polarizations, the gauge invariance can still hold for them. For massive bosons, only when $p^3=0$, the Ward identity is secured and the longitudinal polarization states coincide with each other, $\epsilon^3_R$ = $\epsilon^3_L$ =$(0,0,0,1)$.  In the following, the sums of those polarization vectors are provided
\begin{eqnarray*}
 \sum_{\lambda=1}^{3}\epsilon^{\lambda,\mu}_{X} \epsilon^{\lambda,\nu}_{X}=-g^{\mu\nu}+\frac{p^\mu p^\nu}{p^2}, \,\,&{\rm and}& \,\, \sum_{\lambda=1}^{3}\epsilon^{\lambda,\mu}_{R} \epsilon^{\lambda,\nu}_{L}=\left.-g^{\mu\nu}+\frac{p^\mu p^\nu}{p^2}\right|_{\begin{matrix}\mathsmaller{p_0=0}\\
\,\,\mathsmaller{ {\rm or } \,\,p_3=0}\end{matrix}},
\end{eqnarray*}
where $X=R,L$. The formula of the sum for both of the L- and R-type only applies when $p^3=0$, in the $s$-channel.

\end{document}